\documentclass[12pt]{article}
\usepackage{enumerate}
\usepackage[authoryear]{natbib}
\usepackage{url} 
\usepackage{multirow}
\usepackage{booktabs}
\usepackage{threeparttable}
\usepackage[colorlinks,bookmarksopen,bookmarksnumbered,citecolor=blue,urlcolor=blue,linkcolor=blue]{hyperref}
\usepackage{float}
\usepackage{amsthm,amsmath,enumerate,amsbsy,amsfonts,amssymb,mathabx,amscd,graphicx,algorithm,indentfirst,psfrag,epsf,setspace}
\usepackage{authblk, caption, subcaption}
\usepackage{mathrsfs}
\usepackage{dsfont}
\usepackage{makecell}
\usepackage{array}

\newtheorem{theorem}{Theorem}
\newtheorem{corollary}{Corollary}

\newtheorem{remark}{Remark}
\newtheorem{example}{Example}

\newtheorem{assumption}{Assumption}
\numberwithin{remark}{section}

\newcommand{\bA}{{\bf A}}
\newcommand{\bB}{{\bf B}}

\newcommand{\bI}{{\bf I}}

\newcommand{\bs}{{\bf s}}
\newcommand{\bH}{{\bf H}}
\newcommand{\ba}{{\bf a}}

\newcommand{\bY}{{\bf Y}}
\newcommand{\be}{{\bf e}}
\newcommand{\br}{{\bf r}}
\newcommand{\bR}{{\bf R}}
\newcommand{\bx}{{\bf x}}
\newcommand{\by}{{\bf y}}
\newcommand{\bg}{{\bf g}}

\newcommand{\bphi}{\boldsymbol \phi}

\newcommand{\btheta}{\boldsymbol \theta}
\newcommand{\balpha}{\boldsymbol \alpha}
\newcommand{\bTheta}{\boldsymbol \Theta}

\newcommand{\bbeta}{\boldsymbol \beta}

\newcommand{\bvarphi}{\boldsymbol \varphi}

\newcommand{\bzero}{\boldsymbol 0}
\newcommand{\blambda}{\boldsymbol \lambda}
\newcommand{\bLambda}{\boldsymbol \Lambda}
\newcommand{\bsigma}{\boldsymbol \sigma}
\newcommand{\bOmega}{\boldsymbol \Omega}

\newcommand{\eZ}{\mathbb{Z}}
\newcommand{\eR}{\mathbb{R}}

\newcommand{\eE}{\mathbb{E}}

\newcommand{\eC}{\mathbb{C}}
\newcommand{\eB}{\mathbb{B}}

\newcommand{\cN}{\mathcal{N}}
\newcommand{\cF}{\mathcal{F}}

\newcommand{\cL}{\mathcal{L}}

\newcommand{\cT}{\mathcal{T}}
\newcommand{\cJ}{\mathcal{J}}

\newcommand{\rank}{\mathrm{rank}}

\def\W{{ \mathrm{\scriptscriptstyle W} }}
\def\LM{{ \mathrm{\scriptscriptstyle LM} }}
\def\LR{{ \mathrm{\scriptscriptstyle LR} }}
\def\T{{ \mathrm{\scriptscriptstyle T} }}
\def\dif{\mathinner{\mathrm{d}}\hphantom{\mskip-\thinmuskip}}
\let\daccent\d
\let\d\relax
\newcommand\d{
  \ifmmode
    \dif
  \else
  \@ifnextchar\bgroup
    \daccent
  \fi
}

\newcommand{\blind}{0}

\setcitestyle{aysep={,},yysep={;}}

\begin{document}

\def\spacingset#1{\renewcommand{\baselinestretch}%
{#1}\small\normalsize} \spacingset{1}


\if0\blind
{
  \title{\bf On a new robust method of inference for general time series models}
  \author[1]{\normalsize Zihan Wang}
  \author[2]{Xinghao Qiao}
  \author[1]{Dong Li}
  \author[1,3]{Howell Tong}
  \affil[1]{\it Department of Statistics and Data Science, Tsinghua University, Beijing, China}
  \affil[2]{\it Faculty of Business and Economics, The University of Hong Kong,  Hong Kong}
  \affil[3]{\it Department of Statistics, London School of Economics,  London, U.K.}
    \date{}
  \maketitle
} \fi

\if1\blind
{
    \spacingset{2}
  \bigskip
  \bigskip
  \bigskip
  \begin{center}
    {\LARGE\bf On a new robust method of inference for general time series models}
\end{center}
  \medskip
} \fi

\spacingset{1.3} 

\begin{abstract}
In this article, we propose a novel logistic quasi-maximum likelihood estimation (LQMLE) for general parametric time series models. Compared to the classical Gaussian QMLE and existing robust estimations, it enjoys many distinctive advantages, such as robustness in respect of  distributional misspecification and heavy-tailedness of the innovation, more resiliency to outliers, smoothness and strict concavity of the log logistic quasi-likelihood function, and boundedness of the influence function among others. Under some mild conditions, we establish the strong consistency and asymptotic normality of the LQMLE. Moreover, we propose a new and vital parameter identifiability condition to ensure  desirable asymptotics of the LQMLE. Further, based on the LQMLE, we consider the Wald test and the Lagrange multiplier test for the unknown parameters, and derive the limiting distributions of the corresponding test statistics. The applicability of our methodology is demonstrated by several time series models, including DAR, GARCH, ARMA-GARCH, DTARMACH, and EXPAR. Numerical simulation studies are carried out to assess the finite-sample performance of our methodology, and an empirical example is analyzed to illustrate its usefulness. 
\end{abstract}



\bigskip

\vfill

\newpage
\spacingset{1.7} 

\section{Introduction}
\label{sec.intro}
Time series models play a crucial role across diverse fields, such as data science, economics, finance, engineering, environmental science,  epidemiology,
healthcare, hydrology, and sociology, among others. Researchers and practitioners often need to model time series data 
to reveal temporal dependent structures, as well as to understand dynamic mechanisms and patterns in data, to predict future trends, and to make decisions. See, for example,  \cite{box2015time}, \cite{tsay2010analysis}, 
\cite{WSR2024}, etc. In most practical scenarios, the modeling is based on the 
first two conditional moments. Specifically, a  general parametric time series model
\citep{bardet2009asymptotic, ling2010general}  takes typically the form
\begin{equation}\label{eq.model}
y_t=g(\bY_{t-1},\btheta)+\sigma(\bY_{t-1},\btheta)\eta_t,\quad t\in\eZ,
\end{equation}
where $\bY_{t-1}=(y_{t-1}, y_{t-2}, \dots)^{\T}\in\eR^{\infty}$, or $\bY_{t-1}=(y_{t-1},\dots,y_{t-m})^{\T}\in\eR^m$, $\btheta\in\eR^d$ is the unknown parameter vector of interest, $g(\bY,\btheta):\eR^\infty\ ({\rm or\ }\eR^{m})\times\eR^d\to\eR$ and $\sigma(\bY,\btheta):\eR^\infty\ ({\rm or\ }\eR^{m})\times\eR^d\to\eR^+$ are measurable functions, the innovation sequence $\{\eta_t\}$ is independent and identically distributed (i.i.d.) with zero mean, and $\eta_t$ is independent of $\{y_s: s<t\}$. 

Model (\ref{eq.model}) is quite general,  containing many classical time series models as special cases. Specifically,  when $\sigma(\bY_{t-1},\btheta)$ is 
a fixed positive constant, i.e., $\sigma(\bY_{t-1},\btheta)\equiv \sigma>0$, it includes the best known linear time series model, namely the autoregressive moving average (ARMA) model \citep{BrockwellDavis, box2015time}. 
When $g(\bY_{t-1},\btheta)\equiv 0$, model~\eqref{eq.model} reduces to a general conditionally heteroscedastic time series model
\begin{equation}\label{eq.model_2}
y_t=\sigma(\bY_{t-1},\btheta)\eta_t, \quad t\in\eZ,
\end{equation}
which includes the celebrated autoregressive conditional heteroscedasticity (ARCH)  model \citep{Engle1982} and the generalized ARCH (GARCH) model \citep{Bollerslev1986}. A fairly comprehensive review of GARCH models is available in \cite{francq2019garch}.

In the literature, the Gaussian quasi-maximum likelihood estimation (GQMLE) is widely used for statistical inference of parametric time series models: for example, \cite{bardet2009asymptotic} and \cite{ling2010general} for model (\ref{eq.model}), and \cite{RZ2006} for model (\ref{eq.model_2}).  Apart from these, there are numerous studies on the GQMLE of many specific parametric time series models, including linear and nonlinear ones. For instance, the GQMLE (asymptotically equivalent to the least squares estimator and the Whittle estimator) of ARMA models has been frequently studied; see, e.g., \cite{BrockwellDavis}, \cite{Straumannbook}, \cite{yao2006gaussian}, \cite{boubacar2018diagnostic}, and \cite{wilms2023sparse}, among others. For GARCH models, their GQMLE was studied by \cite{Lumsdaine1996}, \cite{HallYao}, \cite{JensenRahbek},  \cite{francq2004maximum, FZ2012}, \cite{straumann2006quasi}, \cite{jiang2021adaptive}, \cite{yu2024matrix}, etc. \cite{francq2019garch} provided a quite comprehensive review of the GQMLE of various conditionally heteroscedastic time series models. For the GQMLE of ARMA-GARCH models, see  \cite{ling1997fractionally},  \cite{ling2003asymptotic}, \cite{francq2004maximum}, \cite{Ling2007} and the references therein.

Although robust in respect of distributional misspecification of the innovation to some extent, 
the GQMLE often results in a loss of efficiency \citep{francq2011two, fan2014quasi}, particularly for heavy-tailed time series models.\footnote{Heavy-tailed time series data are ubiquitous in the real world. The modeling of heavy-tailed time series data is a long-lasting topic that has been attracting much attention, resulting in 
a growing number of monographs; see, e.g.,  \cite{IIWbook}, \cite{BDMbook}, \cite{PengQi},  \cite{KulikSoulier}, \cite{Nolanbook}, \cite{NWZbook}, and \cite{Leipus}, among others.}
Substantial documented evidence has demonstrated heavy-tailness of the innovation in  
financial returns, thus ruling out Gaussianity \citep{BW1992, Bai2003, li2023maximum}.
To improve the efficiency of the GQMLE, many remedial methods were proposed for some specific parametric time series models, including  the least absolute deviation estimation \citep{li2008least, zhu2015lade,zhu2019statistical, zhang2022lade}, the conditional quantile estimation \citep{li2015quantile,zheng2018hybrid,zhu2018linear, wang2022hybrid,zhu2023quantile}, 
and general $M$-type estimation \citep{mukherjee2020bootstrapping}, etc. 
A selective overview of contemporary robust statistics is available in \cite{loh2025theoretical}. 
However, these robust estimation methods are often limited in the sense of \cite{Huber} due to the non-smoothness of the objective function and the unboundedness of the influence function \citep{wooldridge2020consistency}, and therefore  still incur potential loss of efficiency and poor performance to a certain extent. 

To overcome the shortcomings,  \cite{wooldridge2020consistency} proposed an ingenious logistic QMLE (LQMLE) similar in principle to the GQMLE.  Unfortunately, he  focused exclusively on the linear regression settings, omitting completely the time series settings. Perhaps the omission is due to the recognition of  a serious obstacle to do with the identifiability of the parameters, that is relevant  for time series analysis, but not for linear regression.
In this article, after overcoming the obstacle, we extend the LQMLE to cover the general time series model (\ref{eq.model}) and develop 
a comprehensive procedure based on LQMLE.   As far as we know, this article is the first one
that fully explores the LQMLE in the context of time series analysis.

\begin{figure}[!htbp]
\begin{center}
\includegraphics[width=0.8\linewidth]{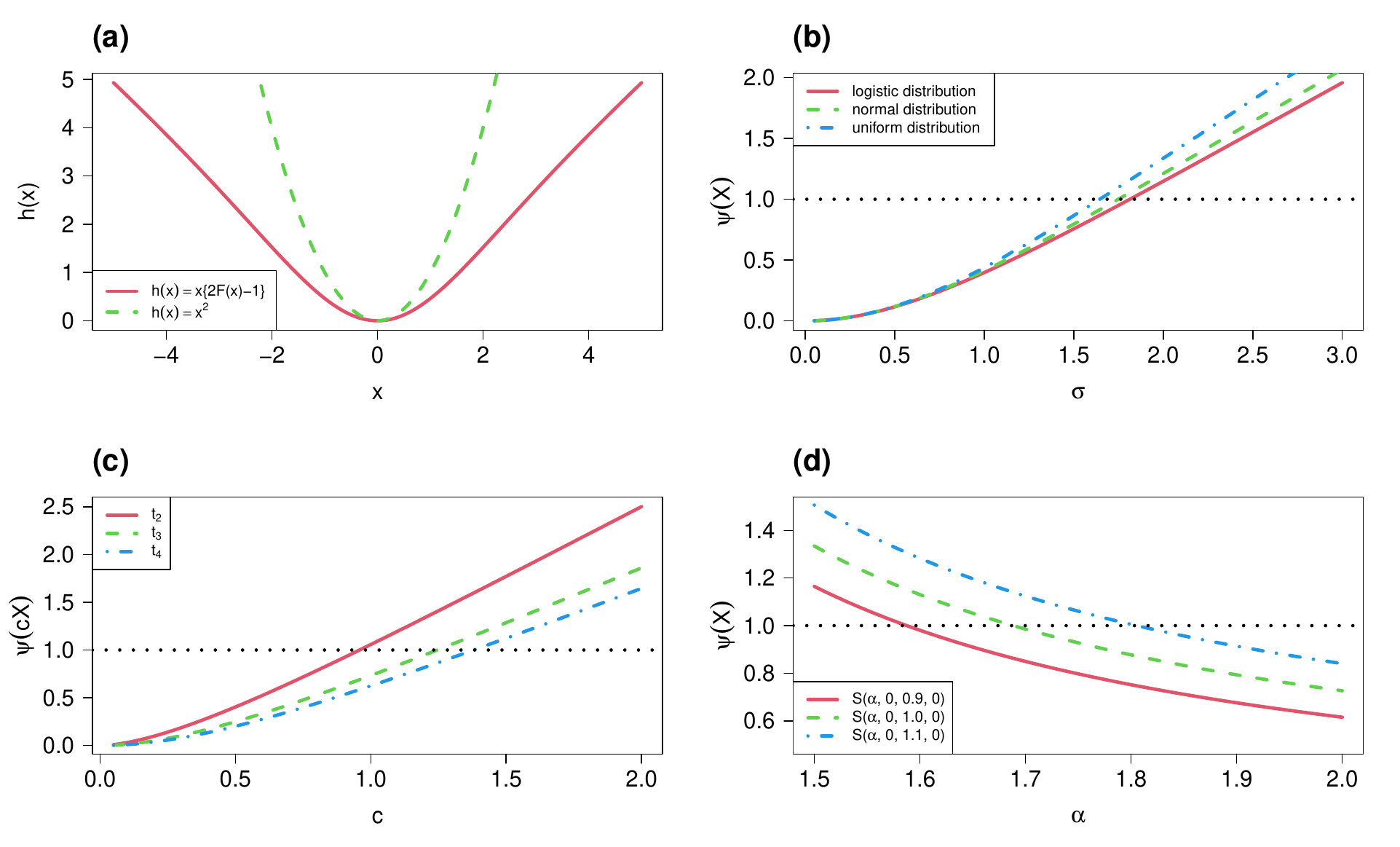}\end{center}
\vspace{-1em}
\caption{\small (a) Plots of $h(x)=x\{2F(x)-1\}$ and $h(x)=x^2$; (b) Plots of $\psi(X)$ for $X\sim\cN(0,\sigma^2),{\rm Logistic}(0,\sqrt{3}\sigma/\pi)$ and $U(-\sqrt{3}\sigma,\sqrt{3}\sigma)$ with different values of $\sigma$; (c) Plots of $\psi(cX)$ for $X\sim t_2$, $t_3$, and $t_4$ with different values of $c>0$; (d) Plots of $\psi(X)$ for $X\sim S(\alpha,0,0.9,0),S(\alpha,0,1,0)$ and $S(\alpha,0,1.1,0)$ with different values of $\alpha$. In (b)-(d), the horizontal dotted lines stand for $\psi(X)=1$.}\label{fig.psi}
\vspace{-1em}
\end{figure}

We shall assume a different moment condition on the innovation rather than the usual condition $\eE\eta_t^2=1$.  
\cite{BerkesHorvath}
made a similar point when studying the non-Gaussian QMLE of GARCH models.
First, 
we define
\begin{equation}\label{eq.psi}
    \psi(\eta_t)=\eE\{h(\eta_t)\}\quad{\rm with}\quad h(x)=x\{2F(x)-1\},\quad x\in\eR,
\end{equation}
where $F(x)=1/\{1+\exp(-x)\}$ is the cumulative distribution function of the standard logistic distribution.
Clearly, $h(x)$ is nonnegative and even; see Figure~\ref{fig.psi}(a). 
We claim
that all parameters are identifiable if $\psi(\eta_t)=1,$ under which the LQMLE is strongly consistent with some extra mild assumptions. 
Lemma~{\color{blue}S.4} of the supplementary material justifies the use  of $\psi(\eta_t)=1$. Such a condition relaxes the restriction of $\eE\eta_t^2<\infty$ to $\eE|\eta_t|<\infty$ since $2F(x)-1\in[-1,1]$ for all $ x\in\eR.$  
Further, it is worth noting that
the conditional variance of $y_t$ (if exists) is proportional to $\sigma^2(\bY_{t-1},\btheta)$, without assuming that  $\eE\eta_t^2=1$. This means that $\sigma^2(\bY_{t-1},\btheta)$ can still  be interpreted as the volatility of models~\eqref{eq.model}-\eqref{eq.model_2}, similar to the interpretation of re-parameterized GARCH models in \cite{fan2014quasi}. 
For an intuitive understanding of $\psi(\eta_t)=1$, we give an example as follows.
\begin{example}\label{ex.psi_1}
Some examples of distributions satisfying $\psi(\eta_t)=1$ include: $\mathrm{(i)}$ the standard logistic distribution ${\rm Logistic}(0,1);$ $\mathrm{(ii)}$ the normal distribution $\cN(0,\sigma^2)$ with $\sigma\approx1.75;$ $\mathrm{(iii)}$ the uniform distribution $U(-a,a)$ with $a\approx2.85;$ $\mathrm{(iv)}$ the Student's $t_{\nu}$-distributions with degrees of freedom $\nu$, e.g., $c\, t_2$ with $c\approx0.96$ and $c\, t_3$ with $c\approx1.25;$ $\mathrm{(v)}$ the standard symmetric stable distribution $S(\alpha,0,1,0)$ with $\alpha\approx1.69.$ 
See Figure~\ref{fig.psi} for more details. 
\end{example}

The main contributions of this article are fourfold.

\textit{First}, we propose a novel LQMLE for general parametric time series model ~\eqref{eq.model}. Under some mild conditions, it is shown that the LQMLE is strongly consistent and asymptotically normal. Compared to the GQMLE and existing robust estimations in the sense of \cite{Huber}, our LQMLE enjoys the following advantages: (i) it is robust in respect of distributional misspecification and heavy-tailedness of the innovation; (ii) it is more resilient to outliers; (iii) the log logistic quasi-likelihood function is smooth and strictly concave; and (iv) the influence function is bounded. These appealing properties simplify computation and statistical inference.

\textit{Second}, we provide a new condition for model parameter identifiability, namely $\psi(\eta_t)=1$, which relaxes the restriction
$\eE\eta_t^2=1$ that is almost routinely used in the GQMLE. The new condition  allows for heavy-tailed distributions such as the $\alpha$-stable distributions with $\alpha\in(1, 2)$ and Student's $t_{\nu}$-distributions with degrees of freedom $\nu\in(1, 2]$. Additionally, for the asymptotic normality of the LQMLE, it suffices to assume a finite second moment of the innovation, equivalent to Assumption \ref{ass.eta_2}. 
This suggests that the innovation following Student's $t_{\nu}$-distribution with $\nu\in(2, 4]$ is also admissible. 

\textit{Third}, based on the LQMLE, we study the Wald test and the Lagrange multiplier test for the unknown parameters, and develop their limiting distributions. Leveraging 
the smoothness and strict concavity of the log logistic quasi-likelihood function, we can construct consistent estimators of asymptotic covariance matrices in the above test statistics with ease. 

\textit{Last}, to illustrate the applicability of the proposed methodology, we verify technical assumptions
for several nonlinear time series models, namely the DAR, GARCH, ARMA-GARCH, DTARMACH, and EXPAR models. 
 Meanwhile, Monte Carlo simulation studies are conducted to examine the performance of our methodology and an empirical example is analyzed.

The remainder of this article is organized as follows. Section~\ref{sec.theory} presents the estimation and inference methodology as well as testing, and establishes related theoretical results. Section~\ref{sec.application} provides concrete applications of our methodology in several specific nonlinear time series models. Section~\ref{sec.simulation} assesses the finite-sample performance of our methodology by Monte Carlo simulation studies. Section~\ref{sec.real} gives an empirical study on treasury yield curve rates. Section~\ref{sec.discussion} concludes. All proofs of main theoretical results as well as additional simulation results are relegated to the supplementary material.

{\bf Notations}. For any vector $\ba,$ we let $\Vert\ba\Vert=\sqrt{\ba^{\T}\ba}$. $\bzero_d$ is a $d$-dimensional vector of zeros. For any matrix $\bA,$ we let $\Vert\bA\Vert=\sqrt{\lambda_{\max}(\bA^{\T}\bA)}$, where $\lambda_{\max}(\cdot)$ denotes the largest eigenvalue of a matrix, and let $\rank(\bA)$ be the rank of $\bA$. $\eR^k$ is the $k$-dimensional Euclidean space ($1\leq k\leq \infty$), and we write $\eR=\eR^1$ and $\eR^+=(0, \infty)$. $\eZ=\{0,\pm1,\pm2,...\}$ and $\eC$ denote the sets of integer and complex numbers, respectively. For $x,y \in {\mathbb R},$ we use $x \wedge y = \min(x,y).$ For two positive sequences $\{a_n\}$ and $\{b_n\}$, we write $a_n\lesssim b_n$ or $a_n=O(b_n)$ or $b_n\gtrsim a_n$ if there exists a positive constant $c$ such that $a_n/b_n \leq c$. We write $a_n \asymp b_n$ if and only if $a_n \lesssim b_n$ and $b_n\lesssim a_n$ hold simultaneously.
The symbols `$\to_p$' and `$\to_d$' mean convergence in probability and convergence in distribution, respectively. $\cF_{t}$ denotes the sigma-algebra generated by random variables $\{y_j: j\leq t\}$ for each $t$, i.e., $\cF_{t}=\sigma(\{y_j: j\leq t\})$.  For a given random variable $X$, $X\in \cF_{t}$ means that $X$ is measurable with respect to (w.r.t.) $\cF_{t}$.

\section{Methodology}\label{sec.theory}
\subsection{Logistic QMLE with Asymptotics}\label{subsec.main}

Let $\btheta$ be the parameter and $\bTheta$ be the parameter space. Suppose that the observations $\{y_1,\dots,y_n\}$ are from model \eqref{eq.model} with the true parameter $\btheta_0.$ 
To handle the initial value problem involved in optimizing the objective function (\ref{obj-fun-L}) below, we first let $\widetilde{\bY}_0=(\widetilde{y}_0,\widetilde{y}_{-1},\dots)^{\T}$ be an initial value and then define $\widetilde{\bY}_{t-1}=(y_{t-1},\dots,y_1, \widetilde{\bY}_0^{\T})^{\T}$ for $t\geq1$. Denote $\widetilde{g}_t(\btheta)=g(\widetilde{\bY}_{t-1},\btheta)$ and $\widetilde{\sigma}_t(\btheta)=\sigma(\widetilde{\bY}_{t-1},\btheta)$ for $t\geq 1$.
The (conditional) log logistic quasi-likelihood function is defined as 
\begin{flalign}\label{obj-fun-L}
\widetilde{\cL}_n(\btheta)=\sum_{t=1}^{n}\widetilde{\ell}_t(\btheta)\quad{\rm with}\quad\widetilde{\ell}_t(\btheta)=-\log\widetilde{\sigma}_t(\btheta)+\log f\left(\frac{y_t-\widetilde{g}_t(\btheta)}{\widetilde{\sigma}_t(\btheta)}\right),
\end{flalign}
where $f(x)$ is the density of the standard logistic distribution, i.e., 
\begin{equation}
    \label{eq.f_density}
    f(x)=\frac{\exp(-x)}{\{1+\exp(-x)\}^2}, \quad x\in\eR.
\end{equation}
The LQMLE of $\btheta_0$ is defined as
\begin{equation}
    \label{eq.lqmle}
    \widehat{\btheta}_n=\arg\max_{\btheta\in\bTheta}\widetilde{\cL}_n(\btheta).
\end{equation}
To facilitate the study on the asymptotic properties of $\widehat{\btheta}_n$, we define the theoretical log logistic 
quasi-likelihood function as
\begin{flalign*}
\cL_n(\btheta)=\sum_{t=1}^{n}\ell_t(\btheta)\quad{\rm with}\quad\ell_t(\btheta)=-\log\sigma_t(\btheta)+\log f\left(\frac{y_t-g_t(\btheta)}{\sigma_t(\btheta)}\right),
\end{flalign*}
where $g_{t}(\btheta)=g(\bY_{t-1},\btheta)$ and $\sigma_{t}(\btheta)=\sigma(\bY_{t-1},\btheta)$ for simplicity.

To obtain the strong consistency of $\widehat{\btheta}_n$, the following assumptions are needed. 

\begin{assumption}
    \label{ass.compact}
    The parameter space $\bTheta$ is compact.
\end{assumption}

\begin{assumption}
    \label{ass.eta}
    $\{\eta_t\}$ is a sequence of i.i.d. symmetric random variables with $\psi(\eta_t)=1$, where $\psi(\cdot)$ is defined in \eqref{eq.psi}.
\end{assumption}

\begin{assumption}
    \label{ass.y}
    $\{y_t\}$ is strictly stationary and ergodic.
\end{assumption}

\begin{assumption}
    \label{ass.lower_bound}
    There exists a constant $\underline{\sigma}>0$ such that $\sigma_t(\btheta)>\underline{\sigma}$ a.s. for any $\btheta\in\bTheta.$
\end{assumption}

\begin{assumption}
    \label{ass.continuous}
    $g(\bY,\btheta)$ and $\sigma(\bY,\btheta)$ are continuous functions w.r.t. $\btheta\in\bTheta.$
\end{assumption}

\begin{assumption}
    \label{ass.identity}
    $(g_t(\btheta),\sigma_t(\btheta))=(g_t(\btheta_0),\sigma_t(\btheta_0))$ a.s. if and only if $\btheta=\btheta_0.$
\end{assumption}

\begin{assumption}
    \label{ass.init_moment}
    $\eE\sup_{\btheta\in\bTheta}|g_t(\btheta)|^{\iota}+\eE\sigma^{\iota}_t(\btheta_0)<\infty$ for some constant $\iota>0.$
\end{assumption}

\begin{assumption}
    \label{ass.init_sigma}
    There exists a nonnegative random variable $C_1\in\cF_{0}$ independent of $\btheta$, and a constant $\rho_1\in(0,1)$ such that $\sup_{\btheta\in\bTheta}\{|\widetilde{g}_t(\btheta)-g_t(\btheta)|+|\widetilde{\sigma}_t^2(\btheta)-\sigma_t^2(\btheta)|\}\le C_1\rho_1^t$ a.s. for all $t\geq 1$.
\end{assumption}

\begin{remark}\label{rmk.ass_consistency}
Assumption~\ref{ass.compact} is standard 
for parametric time series models. In Assumption~\ref{ass.eta}, the identifiability condition $\psi(\eta_t)=1$ implies that $\eE|\eta_t|<\infty$, requiring only a first-order moment condition on $\eta_t$. This relaxes the second-order moment condition necessary for the GQMLE \citep{jeantheau1998strong,ling2003asymptotic}. Together with Assumption~\ref{ass.identity}, 
the restriction $\psi(\eta_t)=1$ serves as the key identifiability condition on $\btheta_0$. 
Additionally, the assumption of symmetry of $\eta_t$ is relatively weak, as it is satisfied by many commonly used random variables. Assumption~\ref{ass.y} is commonly used 
in general nonlinear time series models. 
Assumptions~\ref{ass.lower_bound} and \ref{ass.continuous} are similarly adopted in \cite{jeantheau1998strong}, \cite{francq2015risk}, etc. Assumption~\ref{ass.lower_bound} also implies $\eE\sup_{\btheta\in\bTheta}\{\ell^+_t(\btheta)\}<\infty.$  
Assumption~\ref{ass.init_moment} is mild, requiring only fractional moments on the mean and volatility functions.
Assumption~\ref{ass.init_sigma} is commonly used in existing nonlinear time series analysis literature such as \cite{francq2015risk}, and the exponential decay rates are satisfied by most stationary time series models. Section~\ref{sec.application} below gives some important examples as illustrations. \cite{ling2010general} provided a different initial condition, $\eE\sup_{\btheta\in\bTheta}|\widetilde{\ell}_t(\btheta)-\ell_t(\btheta)|=O(t^{-v})$, for some constant $v>0$ and all $t\geq1$, which can be guaranteed by assuming $\eE\sup_{\btheta\in\bTheta}|g_t(\btheta)|+\eE\sigma_t^2(\btheta_0)<\infty$ and $\eE\sup_{\btheta\in\bTheta}\{|\widetilde{\sigma}_t^2(\btheta)-\sigma_t^2(\btheta)|+|\widetilde{g}_t(\btheta)-g_t(\btheta)|\}=O(t^{-v})$, while we only need $\eE\sup_{\btheta\in\bTheta}|g_t(\btheta)|^{\iota}+\eE\sigma_t^{\iota}(\btheta_0)<\infty$ for some $\iota>0$ in Assumption~\ref{ass.init_moment}, which is weaker than theirs.
\end{remark}

The following theorem states the strong consistency of $\widehat{\btheta}_n$.

\begin{theorem}
    \label{thm.consistency}
    If Assumptions \ref{ass.compact}--\ref{ass.init_sigma} hold, then $\widehat{\btheta}_n\to\btheta_0$ a.s. as $n\to\infty.$ 
\end{theorem}

To obtain the asymptotic distribution of $\widehat{\btheta}_n$, 
we define
{\footnotesize
\begin{equation}
    \label{eq.s_t}
    \bs_t(\btheta)=\frac{\partial\ell_t(\btheta)}{\partial\btheta}=-\frac{\dot{\bsigma}_t^2(\btheta)}{2\sigma_t^2(\btheta)}-\frac{2\dot{\bg}_t(\btheta)\sigma_t(\btheta)+\dot{\bsigma}_t^2(\btheta)(y_t-g_t(\btheta))/\sigma_t(\btheta)}{2\sigma_t^2(\btheta)}\left\{1-2F\left(\frac{y_t-g_t(\btheta)}{\sigma_t(\btheta)}\right)\right\},
\end{equation}}
where $\dot{\bg}(\btheta)=\partial g_t(\btheta)/\partial\btheta$ and $\dot{\bsigma}_t^2(\btheta)=\partial\sigma_t^2(\btheta)/\partial\btheta$,  and define 
{\footnotesize
\begin{equation}\label{eq.H_t}
\begin{aligned}
\bH_t(\btheta)=&-\frac{\partial^2\ell_t(\btheta)}{\partial\btheta\partial\btheta^{\T}}\\
        =&\frac{\dot{\bsigma}_t^2(\btheta)\{\dot{\bsigma}_t^2(\btheta)\}^{\T}-\sigma_t^2(\btheta)\ddot{\bsigma}_t^2(\btheta)}{2\sigma_t^4(\btheta)}\left[\frac{y_t-g_t(\btheta)}{\sigma_t(\btheta)}\left\{2F\left(\frac{y_t-g_t(\btheta)}{\sigma_t(\btheta)}\right)-1\right\}-1\right]\\
        &+\frac{\dot{\bsigma}_t^2(\btheta)\{\dot{\bsigma}_t^2(\btheta)\}^{\T}}{4\sigma_t^4(\btheta)}\left[\frac{y_t-g_t(\btheta)}{\sigma_t(\btheta)}\left\{2F\left(\frac{y_t-g_t(\btheta)}{\sigma_t(\btheta)}\right)-1\right\}+2\left(\frac{y_t-g_t(\btheta)}{\sigma_t(\btheta)}\right)^2f\left(\frac{y_t-g_t(\btheta)}{\sigma_t(\btheta)}\right)\right]\\
        &+\frac{\dot{\bsigma}_t^2(\btheta)\dot{\bg}_t(\btheta)^{\T}+\dot{\bg}_t(\btheta)\{\dot{\bsigma}_t^2(\btheta)\}^{\T}}{2\sigma_t^3(\btheta)}\left\{2F\left(\frac{y_t-g_t(\btheta)}{\sigma_t(\btheta)}\right)-1+2\left(\frac{y_t-g_t(\btheta)}{\sigma_t(\btheta)}\right)f\left(\frac{y_t-g_t(\btheta)}{\sigma_t(\btheta)}\right)\right\}\\
        &+\frac{\ddot{\bg}_t(\btheta)}{\sigma_t(\btheta)}\left\{1-2F\left(\frac{y_t-g_t(\btheta)}{\sigma_t(\btheta)}\right)\right\}+\frac{2\dot{\bg}_t(\btheta)\dot{\bg}_t(\btheta)^{\T}}{\sigma_t^2(\btheta)}f\left(\frac{y_t-g_t(\btheta)}{\sigma_t(\btheta)}\right),
    \end{aligned}
\end{equation}}
where $\ddot{\bsigma}_t^2(\btheta)=\partial^2\sigma_t^2(\btheta)/\partial\btheta\partial\btheta^{\T}$ and $\ddot{\bg}_t(\btheta)=\partial^2g_t(\btheta)/\partial\btheta\partial\btheta^{\T}$. For the reduced model~\eqref{eq.model_2}, $\bs_t(\btheta)$ and $\bH_t(\btheta)$ can be simplified as
\begin{equation}
    \label{eq.H_tilde}
    \begin{aligned}
        \widebar{\bs}_t(\btheta)=&\frac{\dot{\bsigma}_t^2(\btheta)}{2\sigma_t^2(\btheta)}\left[\frac{y_t}{\sigma_t(\btheta)}\left\{2F\left(\frac{y_t}{\sigma_t(\btheta)}\right)-1\right\}-1\right],\\
        \widebar{\bH}_t(\btheta)=&\frac{\dot{\bsigma}_t^2(\btheta)\{\dot{\bsigma}_t^2(\btheta)\}^{\T}-\sigma_t^2(\btheta)\ddot{\bsigma}_t^2(\btheta)}{2\sigma_t^4(\btheta)}\left[\frac{y_t}{\sigma_t(\btheta)}\left\{2F\left(\frac{y_t}{\sigma_t(\btheta)}\right)-1\right\}-1\right]\\
        &+\frac{\dot{\bsigma}_t^2(\btheta)\{\dot{\bsigma}_t^2(\btheta)\}^{\T}}{4\sigma_t^4(\btheta)}\left[\frac{y_t}{\sigma_t(\btheta)}\left\{2F\left(\frac{y_t}{\sigma_t(\btheta)}\right)-1\right\}+2\left(\frac{y_t}{\sigma_t(\btheta)}\right)^2f\left(\frac{y_t}{\sigma_t(\btheta)}\right)\right].
    \end{aligned}
\end{equation}

Let $\widetilde{\bs}_t(\btheta)=\partial\widetilde{\ell}_t(\btheta)/\partial\btheta$ and $\widetilde{\bH}_t(\btheta)=-\partial^2\widetilde{\ell}_t(\btheta)/\partial\btheta\partial\btheta^{\T}$. To obtain the asymptotic distribution of $\widehat{\btheta}_n$, more assumptions are needed. 

\begin{assumption}
    \label{ass.interior}
    The true parameter $\btheta_0$ is an interior point of $\bTheta$.
\end{assumption}

\begin{assumption}
    \label{ass.eta_2}
    $0<\eE\{h(\eta_t)-1\}^2<\infty.$
\end{assumption}

\begin{assumption}
    \label{ass.twice}
    $g(\bY,\btheta)$ and $\sigma(\bY,\btheta)$ are twice continuously differentiable w.r.t. $\btheta\in\bTheta$. 
\end{assumption}

\begin{assumption}
    \label{ass.dot_g}
    $\eE\Vert\dot{\bg}_t(\btheta_0)/\sigma_t(\btheta_0)\Vert<\infty$ and $\eE\Vert\dot{\bsigma}_t^2(\btheta_0)/\sigma_t^2(\btheta_0)\Vert<\infty.$
\end{assumption}

\begin{assumption}
    \label{ass.rank}
    There exist no non-zero vector $\bx\in\eR^d$ such that $\bx^{\T}\dot{\bsigma}_t^2(\btheta_0)=0$ a.s., or there exist no non-zero vector $\by\in\eR^d$ such that $\by^{\T}\dot{\bg}_t(\btheta_0)=0$ a.s.
\end{assumption}

\begin{assumption}
    \label{ass.Ht}
    $\eE\sup_{\btheta\in \eB_{\kappa}(\btheta_0)}\Vert\bH_t(\btheta)\Vert<\infty$ for some $\kappa>0$, where $\eB_{\kappa}(\btheta_0)=\{\btheta:\Vert\btheta-\btheta_0\Vert\le\kappa\}.$
\end{assumption}

\begin{assumption}
    \label{ass.init_s_H}
    $\sup_{\btheta\in\bTheta}\{\Vert\widetilde{\bs}_t(\btheta)-\bs_t(\btheta)\Vert+\Vert\widetilde{\bH}_t(\btheta)-\bH_t(\btheta)\Vert\}\le C_2\rho_2^t$ a.s. for all $t\geq 1$, where $C_2$ and $\rho_2$ are defined similarly to those in Assumption~\ref{ass.init_sigma}. 
\end{assumption}

\begin{remark}\label{rmk.ass_nor}
Assumption~\ref{ass.interior} is standard.
Since $h(x)=x\{2F(x)-1\}\asymp x$ as $x\to\infty$, Assumption~\ref{ass.eta_2} is equivalent to that $\eta_t$ has finite second-order moment, which weakens the finite fourth-order moment condition required for the asymptotic normality of the GQMLE in the literature, allowing innovations to follow the Student's $t_\nu$-distribution with degrees of freedom $\nu\in(2, 4]$. 
Assumption~\ref{ass.twice} ensures that $\bA_0:=\eE\{\bH_t(\btheta_0)\}$ and $\bB_0:=\eE\{\bs_t(\btheta_0)\bs_t(\btheta_0)^{\T}\}$ are well defined. 
Assumption~\ref{ass.dot_g} implies that $\eE\Vert\dot{\bg}_t(\btheta_0)\dot{\bg}_t(\btheta_0)^{\T}/\sigma_t^2(\btheta_0)\Vert<\infty$ and $\eE\Vert\dot{\bsigma}_t^2(\btheta_0)\{\dot{\bsigma}_t^2(\btheta_0)\}^{\T}/\sigma_t^4(\btheta_0)\Vert<\infty$, which, together with Assumption~\ref{ass.eta_2},  ensures that $\Vert\bA_0\Vert<\infty$ and $\Vert\bB_0\Vert<\infty$. Assumption~\ref{ass.rank} ensures that  both $\bA_0$ and $\bB_0$ are positive definite. Assumptions~\ref{ass.twice} and \ref{ass.rank} are also imposed by \cite{francq2015risk}. For the GARCH formulations, Assumption~\ref{ass.rank} reduces to standard assumptions on lag polynomials. Assumption~\ref{ass.Ht} is standard in the literature \citep{ling2010general}. Assumption \ref{ass.init_s_H} is also commonly used in the literature \citep{francq2015risk}, and the decay rates are satisfied for most stationary time series models. Sufficient conditions for Assumptions~\ref{ass.Ht} and \ref{ass.init_s_H} are provided in Sections~{\color{blue}S.3.1} and {\color{blue}S.3.2} respectively of the supplementary material to illustrate their connections with the mean and volatility functions. 
\end{remark}

\begin{remark}
\label{rmk.finite_dim}
It is noteworthy that when the initial value $\widetilde{\bY}_0$ is finite-dimensional, e.g., $\widetilde{\bY}_0=(\widetilde{y}_0,\widetilde{y}_{-1},\dots, \widetilde{y}_{-(m-1)})^{\T}$, which is met by the DAR and EXPAR models studied in Section~\ref{sec.application} below, Assumptions~\ref{ass.init_sigma} and \ref{ass.init_s_H} are satisfied automatically and thus redundant.
\end{remark}

The following theorem and corollary state the asymptotic distributions of $\widehat{\btheta}_n$ defined in \eqref{eq.lqmle} for models~\eqref{eq.model} and \eqref{eq.model_2}, respectively. 

\begin{theorem}
    \label{thm.clt}
    If Assumptions~\ref{ass.compact}--\ref{ass.init_s_H} hold, then 
    $$
    \sqrt{n}(\widehat{\btheta}_n-\btheta_0)\to_d\cN(\bzero_d,\,\bA_0^{-1}\bB_0\bA_0^{-1}),
    $$
    as $n\to\infty,$ where
    {\small
    $$
        \begin{aligned}
        &\bA_0=\eE\{\bH_t(\btheta_0)\}=\left[1+2\eE\{\eta_t^2f(\eta_t)\}\right] \eE\left[\frac{\dot{\bsigma}_t^2(\btheta_0)\{\dot{\bsigma}_t^2(\btheta_0)\}^{\T}}{4\sigma_t^4(\btheta_0)}\right]+2\eE\{f(\eta_t)\}\,\eE\left\{\frac{\dot{\bg}_t(\btheta_0)\dot{\bg}_t(\btheta_0)^{\T}}{\sigma_t^2(\btheta_0)}\right\},\\
        &\bB_0=\eE\{\bs_t(\btheta_0)\bs_t(\btheta_0)^{\T}\}=\eE\{h(\eta_t)-1\}^2 \eE\left[\frac{\dot{\bsigma}_t^2(\btheta_0)\{\dot{\bsigma}_t^2(\btheta_0)\}^{\T}}{4\sigma_t^4(\btheta_0)}\right]+\eE\{2F(\eta_t)-1\}^2\,\eE\left\{\frac{\dot{\bg}_t(\btheta_0)\dot{\bg}_t(\btheta_0)^{\T}}{\sigma_t^2(\btheta_0)}\right\}.
        \end{aligned}
    $$
    }
\end{theorem}

\begin{corollary}
    \label{coro.clt}
    If the conditions of Theorem~\ref{thm.clt} hold, then for model~\eqref{eq.model_2},
    $
    \sqrt{n}(\widehat{\btheta}_n-\btheta_0)\to_d\cN(\bzero_d,\,4\tau\bOmega_0^{-1})
    $
    as $n\to\infty$, where 
    $
    \tau=\frac{\eE\{h(\eta_t)-1\}^2}{[1+2\eE\{\eta_t^2f(\eta_t)\}]^2}$ and $\bOmega_0=\eE\left[\frac{\dot{\bsigma}_t^2(\btheta_0)\{\dot{\bsigma}_t^2(\btheta_0)\}^{\T}}{\sigma_t^4(\btheta_0)}\right].
    $
\end{corollary}

In empirical applications, to make statistical inference on $\btheta_0$, we usually need to estimate the asymptotic covariance matrix
$\bA_0^{-1}\bB_0\bA_0^{-1}$. To this end, it suffices to construct weakly consistent estimators of $\bA_0$ and $\bB_0$, respectively.  Let
\begin{equation}\label{eq.ASD}
\widehat{\bA}_n=\frac{1}{n}\sum_{t=1}^{n}\widetilde{\bH}_t(\widehat{\btheta}_n)\quad{\rm and }\quad\widehat{\bB}_n=\frac{1}{n}\sum_{t=1}^{n}\widetilde{\bs}_t(\widehat{\btheta}_n)\widetilde{\bs}_t(\widehat{\btheta}_n)^{\T}.
\end{equation}

\begin{theorem}
    \label{thm.inference}
    If Assumptions~\ref{ass.compact}--\ref{ass.init_s_H} hold, then $\widehat{\bA}_n=\bA_0+o_p(1)$ and $\widehat{\bB}_n=\bB_0+o_p(1)$.
\end{theorem}

In empirical applications,  we are often interested in the statistical significance of some specific component of $\btheta_0$ in modeling, i.e., testing the hypothesis: $H_{0j}:\theta_{0j}=0$ v.s. $H_{1j}:\theta_{0j}\neq 0$ for some $j\in\{1,\dots,d\}.$ Then the Student's $t$-type of test statistic can be constructed in the form $\cT_{nj}=\sqrt{n}\big(\be_j^{\T}\widehat{\bA}_n^{-1}\widehat{\bB}_n\widehat{\bA}_n^{-1}\be_j\big)^{-1/2}\widehat{\theta}_{nj},$ where the $j$-th entry of $\be_j$ takes 1 and the rest 0. By Theorems~\ref{thm.clt} and \ref{thm.inference}, under $H_{0j}$, $\cT_{nj}\to_d\cN(0,1)$ as $n\to\infty$. Thus, for a given significance level $\alpha\in(0, 1)$, we can reject $H_{0j}$ if $|\cT_{nj}|>\Phi^{-1}(1-\alpha/2)$, where $\Phi^{-1}(\cdot)$ is the quantile function of the standard normal distribution.

\subsection{Testing}\label{subsec.inference}
Consider the hypothesis
\begin{equation}\label{eq.hypothesis}
H_0:\bR\btheta_0=\br\quad {\rm v.s.\quad } H_1:\bR\btheta_0\neq\br,
\end{equation}
where $\bR$ is a $q\times d$ matrix with full row rank, and $\br$ is a $q$-dimensional vector. 

Based on the LQMLE in Section \ref{subsec.main}, we develop the Wald test and the Lagrange multiplier test for the hypothesis~\eqref{eq.hypothesis}. Under $H_0$, by Theorems~\ref{thm.clt} and \ref{thm.inference}, it follows that
\begin{equation}\label{eq.limit_wald}\big(\bR\widehat{\bA}_n^{-1}\widehat{\bB}_n\widehat{\bA}_n^{-1}\bR^{\T}\big)^{-1/2}\sqrt{n}(\bR\widehat{\btheta}_n-\br)\to_d\cN(\bzero_q,\,\bI_q).
\end{equation}
Then, the Wald test statistic is the inner product of the left-hand side of \eqref{eq.limit_wald}, that is, 
\begin{equation}
    \label{eq.test_wald}
    \cT_n^{\W}=n(\bR\widehat{\btheta}_n-\br)^{\T}\big(\bR\widehat{\bA}_n^{-1}\widehat{\bB}_n\widehat{\bA}_n^{-1}\bR^{\T}\big)^{-1}(\bR\widehat{\btheta}_n-\br).
\end{equation}

For the Lagrange multiplier test, we consider the maximization problem \eqref{eq.lqmle}  subject to the constraint $\bR\btheta=\br.$ The constrained LQMLE $\widetilde{\btheta}_n$ of $\btheta_0$ is defined as 
\begin{equation}
    \label{eq.lqmle_con}
    \widetilde{\btheta}_n=\arg\max_{\btheta\in\bTheta}\widetilde{\cL}_n(\btheta),\quad {\rm s.t.\quad }\bR\btheta=\br.
\end{equation}
The Lagrangian of this constrained problem is defined as $\cJ_n(\btheta,\blambda)=\widetilde{\cL}_{n}(\btheta)+n\blambda^{\T}(\bR\btheta-\br),$ where $\blambda$ is the vector of Lagrange multipliers. By the Lagrange duality theory, there exists $\widetilde{\blambda}_n\in\eR^q$ such that the solution of \eqref{eq.lqmle_con} is exactly the solution to the unconstrained problem $\arg\max_{\btheta\in\bTheta}\cJ_n(\btheta,\widetilde{\blambda}_n).$ Define $\bLambda=(\bR\bA_0^{-1}\bR^{\T})^{-1}\bR\bA_0^{-1}\bB_0\bA_0^{-1}\bR^{\T}(\bR\bA_0^{-1}\bR^{\T})^{-1}.$ It can be shown that $\bLambda^{-1/2}\sqrt{n}\widetilde{\blambda}_n\to_d\cN(\bzero_q,\,\bI_q).$ Once we obtain the constrained LQMLE $\widetilde{\btheta}_n$, a plug-in estimator of $\bLambda$ follows, namely $\widetilde{\bLambda}_n=(\bR\widetilde{\bA}_n^{-1}\bR^{\T})^{-1}\bR\widetilde{\bA}_n^{-1}\widetilde{\bB}_n\widetilde{\bA}_n^{-1}\bR^{\T}(\bR\widetilde{\bA}_n^{-1}\bR^{\T})^{-1},$ where $\widetilde{\bA}_n=n^{-1}\sum_{t=1}^{n}\widetilde{\bH}_t(\widetilde{\btheta}_n)$ and $\widetilde{\bB}_n=n^{-1}\sum_{t=1}^{n}\widetilde{\bs}_t(\widetilde{\btheta}_n)\widetilde{\bs}_t(\widetilde{\btheta}_n)^{\T}.$ Then, the Lagrange multiplier test statistic can be defined as 
\begin{equation}\label{eq.test_LM}\cT_n^{\LM}=n\widetilde{\blambda}_n^{\T}\widetilde{\bLambda}_n^{-1}\widetilde{\blambda}_n.
\end{equation}
The limiting distributions of the Wald test statistic and the Lagrange multiplier test statistic follow from the asymptotics of $\widehat{\btheta}_n$ and $\widetilde{\btheta}_n$ and the continuous mapping theorem.

\begin{theorem}
    \label{thm.test}
    If Assumptions~\ref{ass.compact}--\ref{ass.init_s_H} hold, then under $H_0$: $\mathrm{(i)}$ $\cT_n^{\W}\to_d\chi^2(q)$ as $n\to\infty$, and $\mathrm{(ii)}$ $\cT_n^{\LM}\to_d\chi^2(q)$ as $n\to\infty$, where $\cT_n^{\W}$ and $\cT_n^{\LM}$ are defined in \eqref{eq.test_wald} and \eqref{eq.test_LM}, respectively, and $\chi^2(q)$ is a chi-squared distribution with degrees of freedom $q$ with $q=\rank(\bR)$.
\end{theorem}

\begin{remark}
    Another classical large sample test is the likelihood ratio test, which compares the performance of the constrained and unconstrained specifications. In the hypothesis~\eqref{eq.hypothesis}, the (quasi-) likelihood ratio test statistic is defined as $\cT_n^{\LR}=2\{\widetilde{\cL}_n(\widetilde{\btheta}_n)-\widetilde{\cL}_n(\widehat{\btheta}_n)\}.$ The limiting distribution of $\cT_n^{\LR}$ relies on the information matrix equality $\bA_0=\bB_0$, however, which is generally not satisfied for the cases of distributional misspecifications of innovations. 
\end{remark}

\section{Applications}
\label{sec.application}

In this section, we apply the proposed method and theory to five important time series models in nonlinear time series analysis. We focus on checking relevant technical assumptions used to ensure the asymptotics of the LQMLE. 

\subsection{DAR Model}\label{subsec.DAR}
A time series $\{y_t\}$ is said to follow
a double autoregressive (DAR) model of order $(p, q)$ if it satisfies the stochastic recurrence equation:
\begin{equation}\label{eq.DAR}
    y_t=\phi_0+\sum_{i=1}^{p}\phi_iy_{t-i}+\eta_t\Big(\alpha_0+\sum_{j=1}^{q}\alpha_jy_{t-j}^2\Big)^{1/2},\quad t\in\eZ,
\end{equation}
where $\phi_i\in\eR$, $\alpha_0>0$, $\alpha_j\ge0$ for $1\leq j\leq q$, $\phi_p\alpha_q\neq0$, $\{\eta_t\}$ is a sequence of i.i.d. random variables with $\eta_t$  independent of $\cF_{t-1}$. For model \eqref{eq.DAR} with symmetric innovations, \cite{guegan1994probabilistic} proved it is strictly stationary and ergodic if the top Lyapunov exponent $\gamma<0$. Particularly, for $p=q=1$, it is shown that $\gamma=\eE\log|\phi_1+\eta_t\sqrt{\alpha_1}|$. By Theorem 2.1 in \cite{cline2004stability}, a sufficient condition for strict stationarity and ergodicity of $\{y_t\}$ is that $\sum_{i=1}^{p\vee q}(|\phi_i|^2+\alpha_i\eE|\eta_t|)<1$ with $\phi_i=0$ for $i>p$ and $\alpha_i=0$ for $i>q,$ and $\eta_t$ having a continuous and positive density over $\eR$. 

Let $d=p+q+2$, $m=\max\{p, q\}$, $\btheta=(\bphi^{\T},\balpha^{\T})^{\T}$ with $\bphi=(\phi_0,\dots,\phi_p)^{\T}$ and $\balpha=(\alpha_0,\dots,\alpha_q)^{\T}$, $\bY_t=(y_t,\dots,y_{t-m+1})^{\T},g_t(\btheta)=\phi_0+\sum_{i=1}^{p}\phi_iy_{t-i}$, and $\sigma_t(\btheta)=(\alpha_0+\sum_{j=1}^{q}\alpha_jy_{t-j}^2)^{1/2}.$ 
Assumption~\ref{ass.y} is guaranteed by $\gamma<0$. Assumption~\ref{ass.lower_bound} is satisfied due to the compactness of $\bTheta$.  Assumptions~\ref{ass.continuous}, \ref{ass.identity}, \ref{ass.interior}, \ref{ass.twice}, and \ref{ass.rank} hold automatically. Assumption~\ref{ass.init_moment} is satisfied by assuming that $\eE|y_t|^{\iota}<\infty$ for some $\iota>0$.  
In addition, note that $\dot{\bg}_t(\btheta)=(1,y_{t-1},\dots,y_{t-p},\bzero_{q+1}^{\T})^{\T}$, and $\dot{\bsigma}_t^2(\btheta)=(\bzero_{p+1}^{\T},1,y_{t-1}^2,\dots,y_{t-q}^2)$. Then, Assumptions~\ref{ass.dot_g} and \ref{ass.Ht} are met by the compactness of $\bTheta$ and $\eE y_t^2<\infty$, which is implied by Assumption~\ref{ass.eta_2} and is not necessary if $q\ge p.$ Note that the dimension of the initial value is finite for model \eqref{eq.DAR}. 
Thus, together with Remark~\ref{rmk.finite_dim}, the proposed LQMLE for the DAR model~\eqref{eq.DAR} is strongly consistent and asymptotically normal.

\subsection{GARCH Model}
\label{subsec.GARCH}
Consider the classical GARCH$(p, q)$ model
\begin{equation}
    \label{eq.GARCH}
y_t=\sigma_t\eta_t\quad{\rm with}\quad \sigma_t^2=\alpha_0+\sum_{i=1}^{p}\alpha_iy_{t-i}^2+\sum_{j=1}^{q}\beta_j\sigma_{t-j}^2,\quad t\in\eZ,
\end{equation}
where $\alpha_0>0$, $\alpha_i\ge0$, $\beta_j\ge0$ for $i,j\geq 1$, $\alpha_p\beta_q>0$, $\{\eta_t\}$ is a sequence of i.i.d. random variables and $\eta_t$ is independent of $\cF_{t-1}$. By \cite{bougerol1992stationarity}, $\{y_t\}$  is strictly stationary and ergodic if and only if the top Lyapunov exponent $\gamma<0.$ Particularly, when $p=q=1$, it is known that $\gamma=\eE\log(\beta_1+\alpha_1\eta_t^2)$.

Let $d=p+q+1$, $m=\infty$, $\btheta=(\balpha^{\T},\bbeta^{\T})^{\T}$ with $\balpha=(\alpha_0,\dots,\alpha_p)^{\T}$ and $\bbeta=(\beta_1,\dots,\beta_q)^{\T}$, $\bY_t=(y_{t},y_{t-1},\dots)^{\T}\in\eR^{\infty}$, $g_t(\btheta)\equiv0,$ and $\sigma_t(\btheta)=\big\{\alpha_0+\sum_{i=1}^{p}\alpha_iy_{t-i}^2+\sum_{j=1}^{q}\beta_j\sigma_{t-j}^2(\btheta)\big\}^{1/2}.$ On assuming that $\alpha_i$'s and $\beta_j$'s are bounded away from zero and $\gamma<0$, Assumptions~\ref{ass.compact}--\ref{ass.init_moment} and \ref{ass.interior}--\ref{ass.Ht} can be verified by arguments similar  to Section~\ref{subsec.DAR}. Since the dimension of the initial value $\widetilde{\bY}_0$ in optimizing the objective function \eqref{obj-fun-L} for model~\eqref{eq.GARCH} is infinite, we focus on checking the initial conditions. For simplicity, we here consider a GARCH(1,1) model as an example, noting that  the discussion on higher-order cases is similar. We know that $\beta_1\in(0,1)$ since $\gamma<0.$  Let  $\sigma_t^2(\btheta)=\alpha_0+\alpha_1y_{t-1}^2+\beta_1\sigma_{t-1}^2(\btheta)$ with $\btheta=(\alpha_0,\alpha_1,\beta_1)^{\T}$. By iterations, we have $\sigma_t^2(\btheta)=\alpha_0\sum_{k=1}^{\infty}\beta_1^{k-1}+\alpha_1\sum_{h=1}^{\infty}\beta_1^{h-1}y_{t-h}^2$. Without loss of generality, suppose that the initial value is $\widetilde{\bY}_0=\bf0$, which is most commonly used in optimizing the objective function. Then $\widetilde{\sigma}_t^2(\btheta)=\alpha_0\sum_{k=1}^{\infty}\beta_1^{k-1}+\alpha_1\sum_{h=1}^{t-1}\beta_1^{h-1}y_{t-h}^2$. Thus 
$$
\sigma_t^2(\btheta)-\widetilde{\sigma}_t^2(\btheta)=\alpha_1\sum_{h=t}^{\infty}\beta_1^{h-1}y_{t-h}^2=\alpha_1\sum_{h=0}^{\infty}\beta_1^{t+h-1}y_{-h}^2=\beta_1^t\left(\alpha_1\sum_{h=0}^{\infty}\beta_1^{h-1}y_{-h}^2\right).
$$
Let $C_1(\btheta)=\alpha_1\sum_{h=0}^{\infty}\beta_1^{h-1}y_{-h}^2$ and $C_1=\sup_{\btheta\in\bTheta}|C_1(\btheta)|=\bar{\alpha}_1\sum_{h=0}^{\infty}\bar{\beta}_1^{h-1}y_{-h}^2$, where $\bar{\alpha}_1=\sup\{\alpha_1|\btheta\in\bTheta\}<\infty$ and 
$\bar{\beta}_1=\sup\{\beta_1|\btheta\in\bTheta\}<1$ due to the compactness of $\bTheta$ and $\gamma<0$. Clearly, $C_1\in \cF_{0}$ is a nonnegative random variable independent of $\btheta$.
Thus, Assumption~\ref{ass.init_sigma} is satisfied by letting $\rho_1=\bar{\beta}_1$.
Then, by simple algebraic calculations, we have
$$
\frac{\partial\sigma_t^2(\btheta)}{\partial\btheta}-\frac{\partial\widetilde{\sigma}_t^2(\btheta)}{\partial\btheta}=\Big(0,\,\sum_{h=0}^{\infty}\beta_1^{t+h-1}y_{-h}^2,\,\sum_{h=0}^{\infty}\alpha_1(t+h-1)\beta_1^{t+h-2}y_{-h}^2\Big)^{\T},
$$
and thus $\big\Vert\partial\sigma_t^2(\btheta)/\partial\btheta-\partial\widetilde{\sigma}_t^2(\btheta)/\partial\btheta\big\Vert\le \beta_1^t\big[\sum_{h=0}^{\infty}\{\beta_1+(t+h-1)\alpha_1\}\beta_1^{h-2}y_{-h}^2\big]$. Similarly, it can be shown that $\big\Vert\partial^2\sigma_t^2(\btheta)/\partial\btheta\partial\btheta^{\T}-\partial^2\widetilde{\sigma}_t^2(\btheta)/\partial\btheta\partial\btheta^{\T}\big\Vert\le \beta_1^t\big[\sum_{h=0}^{\infty}\{2(t+h-1)+(t+h-1)(t+h-2)\alpha_1/\beta_1\}\beta_1^{h-2}y_{-h}^2\big]$. Note that there exists a constant $\rho_2$ with $0<\rho_2<1$ such that $t^2\bar{\beta}_1^t<\rho_2^t$ for $t$ large enough. Then Assumption~\ref{ass.init_s_H} is satisfied by adopting the sufficient condition discussed in Section~{\color{blue}S.3.1} of the supplementary material. Thus, the proposed LQMLE for the GARCH model~\eqref{eq.GARCH} is strongly consistent and asymptotically normal.

\subsection{ARMA-GARCH Model}
\label{subsec.ARMA-GARCH}
Consider an ARMA$(p,q)$-GARCH$(r,s)$ model
\begin{equation}\label{eq.ARMA-GARCH}
    y_t=\phi_0+\sum_{i=1}^{p}\phi_iy_{t-i}+\sum_{j=1}^{q}\varphi_j\varepsilon_{t-j}+\varepsilon_t \ {\rm with}\ \varepsilon_t=\sigma_t\eta_t\ {\rm and}\ \sigma_t^2=\alpha_0+\sum_{k=1}^{r}\alpha_k\varepsilon_{t-k}^2+\sum_{l=1}^{s}\beta_l\sigma_{t-l}^2,
\end{equation}
where $\phi_i\in\eR$, $\varphi_j\in\eR$, $\phi_p\varphi_q\neq0$, $\alpha_0>0$, $\alpha_k\ge0$, $\beta_l\ge0$, $\alpha_r\beta_s>0$, $\{\eta_t\}$ is i.i.d. and $\eta_t$ is independent of  $\cF_{t-1}$. The strict stationarity and ergodicity of the ARMA-GARCH model were studied in \cite{ling1997fractionally}, \cite{ling2003asymptotic}, etc. 

For illustration, model~\eqref{eq.ARMA-GARCH} can be written in the form of model~\eqref{eq.model}. To this end, we let $\phi(z)=1-\phi_1z-\dots-\phi_pz^p$
and $\varphi(z)=1+\varphi_1z+\dots+\varphi_qz^q$ be the $p$-th and $q$-th degree characteristic polynomials for any $z\in\eC$, respectively.  Suppose that (i)  $\phi(z)\neq 0$, $\varphi(z)\neq 0$ for all $|z|\leq 1$, and $\phi(z)$ and $\varphi(z)$ have no common factors; and (ii) $\sum_{k=1}^{r}\alpha_k+\sum_{l=1}^{s}\beta_l<1$.
Then, under (ii), $\{\sigma_t\}$ is stationary (strictly and weakly) and ergodic. Further, $\{y_t\}$ is (causal) stationary, ergodic, and invertible under (i). Thus, 
by iterations, we can see that 
$ \varepsilon_t=y_t-\phi_0-\sum_{i=1}^{p}\phi_iy_{t-i}-\sum_{j=1}^{q}\varphi_j\varepsilon_{t-j}\in\cF_{t}$. Then the conditional mean function
$\phi_0+\sum_{i=1}^{p}\phi_iy_{t-i}+\sum_{j=1}^{q}\varphi_j\varepsilon_{t-j}\in\cF_{t-1}$, i.e., there exists a measurable function $g_t(\btheta)\in\cF_{t-1}$ such that
$g_t(\btheta)=\phi_0+\sum_{i=1}^{p}\phi_iy_{t-i}+\sum_{j=1}^{q}\varphi_j\varepsilon_{t-j}$.
Similarly, $\sigma_t^2\in\cF_{t-1}$ by noting that $\sigma_t^2$ is a measurable function of $\{\varepsilon^2_{t-j}: j\geq 1\}$. Due to 
$\sigma_t>0$ a.s., we have $\sigma_t\in\cF_{t-1}$. Thus, there exists 
$\sigma_t(\btheta)\in\cF_{t-1}$ such that 
$\sigma_t(\btheta)=\big(\alpha_0+\sum_{k=1}^{r}\alpha_k\varepsilon_{t-k}^2+\sum_{l=1}^{s}\beta_l\sigma_{t-l}^2\big)^{1/2}$.
Therefore, model~\eqref{eq.ARMA-GARCH} has the form of model~\eqref{eq.model}.

For simplicity, we here consider an ARMA(1,1)-GARCH(1,1) model as a benchmark. 
Specifically, $y_t=\phi_0+\phi_1y_{t-1}+\varphi_1\varepsilon_{t-1}+\varepsilon_t$ with $\varepsilon_t=\sigma_t\eta_t$ and $\sigma_t^2=\alpha_0+\alpha_1\varepsilon_{t-1}^2+\beta_1\sigma_{t-1}^2.$ Let $\btheta=(\phi_0,\phi_1,\varphi_1,\alpha_0,\alpha_1,\beta_1)^{\T}.$ Suppose that (i) $^{\prime}$\, $|\phi_1|<1$, $|\varphi_1|<1$, and $\phi_1+\varphi_1\neq0$; 
(ii)$^{\prime}$\, $\alpha_1+\beta_1<1$. By Lemma 2.2 in  \cite{francq2019garch} and the fact $\eE|\eta_t|<\infty$ implied by Assumption~\ref{ass.eta}, there exists some constant $\iota>0$ such that $\eE\sup_{\btheta\in\bTheta}\sigma_t^{\iota}(\btheta)<\infty.$ By iterations, it follows that
\begin{equation}
    \label{eq.iter_ARMA_GARCH}
    y_t=\phi_0\sum_{k=1}^{\infty}(-\varphi_1)^{k-1}+(\phi_1+\varphi_1)\sum_{h=1}^{\infty}(-\varphi_1)^{h-1}y_{t-h}+\varepsilon_t.
\end{equation}
Let $g_t(\btheta)=\phi_0\sum_{k=1}^{\infty}(-\varphi_1)^{k-1}+(\phi_1+\varphi_1)\sum_{h=1}^{\infty}(-\varphi_1)^{h-1}y_{t-h}\in\cF_{t-1}$ and $\varepsilon_t(\btheta)=y_t-g_t(\btheta)\in\cF_{t}$. 
Similar to Section~\ref{subsec.GARCH}, we can get that 
\begin{equation}\label{eq.iter_2}\sigma_t^2(\btheta)=\alpha_0\sum_{l=1}^{\infty}\beta_1^{l-1}+\alpha_1\sum_{b=1}^{\infty}\beta_1^{b-1}\varepsilon_{t-b}^2(\btheta)\in\cF_{t-1}.
\end{equation}
Next, we focus on verifying Assumptions~\ref{ass.init_sigma}, \ref{ass.dot_g}, \ref{ass.Ht}, and \ref{ass.init_s_H}, while the others are straightforward to check. By tedious algebraic calculations, we have 
{\small
$$
\begin{aligned}
    \dot{\bg}_t(\btheta)=&\Big(\sum_{k=1}^{\infty}(-\varphi_1)^{k-1},\sum_{h=1}^{\infty}(-\varphi_1)^{h-1}y_{t-h},\sum_{h=1}^{\infty}[\{(1-h)\phi_1-h\varphi_1\}y_{t-h}-(h-1)\phi_0](-\varphi_1)^{h-2},0,0,0\Big)^{\T},\\
    \dot{\bsigma}_t^2(\btheta)=&\Big(0,0,0,\sum_{l=1}^{\infty}\beta_1^{l-1},\sum_{b=1}^{\infty}\beta_1^{b-1}\varepsilon_{t-b}^2(\btheta),\sum_{l=1}^{\infty}\{(l-1)\alpha_0+(l-1)\alpha_1\varepsilon_{t-l}^2(\btheta)\}\beta_1^{l-2}\Big)^{\T}\\
    &-2\alpha_1\sum_{b=1}^{\infty}\beta_1^{b-1}\varepsilon_{t-b}(\btheta)\dot{\bg}_{t-b}(\btheta),
\end{aligned}
$$}
Then, Assumption~\ref{ass.dot_g} is satisfied by noting that the conditions (i)$^{\prime}$--(ii)$^{\prime}$, the compactness of $\bTheta$, and the fact $\eE|y_t|<\infty$ implied by Assumption~\ref{ass.eta}. Similarly, Assumption~\ref{ass.Ht} is also satisfied by noting that $\eE y_t^2<\infty$ implied by Assumption~\ref{ass.eta_2} and the conditions (i)$^{\prime}$--(ii)$^{\prime}$.

Without loss of generality, suppose that the initial value is $\widetilde{\bY}_0=\bzero.$ Then let $\widetilde{g}_t(\btheta)=g(\widetilde{\bY}_{t-1},\btheta),$ $\widetilde{\sigma}_t(\btheta)=\sigma(\widetilde{\bY}_{t-1},\btheta)$ and $\widetilde{\varepsilon}_t(\btheta)=\widetilde{y}_t-\widetilde{g}_t(\btheta)$. It can be shown that
{\small
$$
\begin{aligned}
    g_t(\btheta)-\widetilde{g}_t(\btheta)=&(\phi_1+\varphi_1)\sum_{h=t}^{\infty}(-\varphi_1)^{h-1}y_{t-h}=\varphi_1^t\Big\{(\phi_1+\varphi_1)\sum_{h=0}^{\infty}(-1)^{t+h-1}\varphi_1^{h-1}y_{-h}\Big\},\\
    \sigma_t^2(\btheta)-\widetilde{\sigma}_t^2(\btheta)=&\alpha_1\sum_{b=0}^{\infty}\beta_1^{t+b-1}\{\varepsilon_{-b}^2(\btheta)-\widetilde{\varepsilon}_{-b}^2(\btheta)\}=\alpha_1\sum_{b=0}^{\infty}\beta_1^{t+b-1}\{\varepsilon_{-b}(\btheta)+\widetilde{\varepsilon}_{-b}(\btheta)\}\{\varepsilon_{-b}(\btheta)-\widetilde{\varepsilon}_{-b}(\btheta)\}\\
    =&\beta_1^t\left[\alpha_1\sum_{b=0}^{\infty}\beta_1^{b-1}\{\varepsilon_{-b}(\btheta)+\widetilde{\varepsilon}_{-b}(\btheta)\}\Big\{y_{-b}+\varphi_1^{-b}(\phi_1+\varphi_1)\big(\sum_{h=0}^{\infty}(-1)^{-b+h-1}\varphi_1^{h-1}y_{-h}\big)\Big\}\right],
\end{aligned}
$$}
which implies Assumption~\ref{ass.init_sigma} by similar arguments in Section~\ref{subsec.GARCH}. By tedious algebraic calculations, it follows that
{\small
$$
\begin{aligned}
    \frac{\partial g_t(\btheta)}{\partial\btheta}-\frac{\partial \widetilde{g}_t(\btheta)}{\partial\btheta}=&\varphi_1^t\Big(0,\sum_{h=0}^{\infty}(-1)^{t+h-1}\varphi_1^{h-1}y_{-h},\sum_{h=0}^{\infty}(-1)^{t+h-2}\varphi_1^{h-2}\{(1-h-t)\phi_1-(t+h)\varphi_1\}y_{-h},0,0,0\Big),\\
    \frac{\partial \sigma_t^2(\btheta)}{\partial\btheta}-\frac{\partial \widetilde{\sigma}_t^2(\btheta)}{\partial\btheta}=&\beta_1^t\Big(0,0,0,0,\sum_{b=0}^{\infty}\beta_1^{b-1}\{\varepsilon_{-b}^2(\btheta)-\widetilde{\varepsilon}_{-b}^2(\btheta)\},\alpha_1\sum_{l=0}^{\infty}(t+l-1)\beta_1^{l-2}\{\varepsilon_{-l}^2(\btheta)-\widetilde{\varepsilon}_{-l}^2(\btheta)\}\Big)\\
    &+\beta_1^t\Big[-2\alpha_1\sum_{b=0}^{\infty}\beta_1^{b-1}\Big\{\varepsilon_{-b}(\btheta)\frac{\partial g_{-b}(\btheta)}{\partial\btheta}-\widetilde{\varepsilon}_{-b}(\btheta)\frac{\partial \widetilde{g}_{-b}(\btheta)}{\partial\btheta}\Big\}\Big].
\end{aligned}
$$}
Similarly, we can also obtain that the expressions of $\partial^2 g_t(\btheta)/\partial\btheta\partial\btheta^{\T}-\partial^2 \widetilde{g}_t(\btheta)/\partial\btheta\partial\btheta^{\T}$ and $\partial^2 \sigma_t^2(\btheta)/\partial\btheta\partial\btheta^{\T}-\partial^2 \widetilde{\sigma}_t^2(\btheta)/\partial\btheta\partial\btheta^{\T}$.
Then Assumption~\ref{ass.init_s_H} is satisfied by the sufficient condition discussed in Section~{\color{blue}S.3.1} of the supplementary material. Thus, the proposed LQMLE for the ARMA-GARCH model~\eqref{eq.ARMA-GARCH} is strongly consistent and asymptotically normal. 

\subsection{DTARMACH Model}
Consider a double threshold ARMA conditional heteroscedasticity (DTARMACH) model, proposed by \cite{ling1999probabilistic} and defined as
\begin{equation}
    \label{eq.DTARMACH}
    \begin{aligned}
        y_t=&\phi_0^{(u)}+\sum_{i=1}^{p}\phi_i^{(u)}y_{t-i}+\sum_{j=1}^{q}\varphi_j^{(u)}\varepsilon_{t-j}+\varepsilon_t,\quad a_{u-1}<y_{t-b}\le a_u,\\
        \varepsilon_t=&\sigma_t\eta_t,\\
        \sigma_t^2=&\alpha_0^{(v)}+\sum_{k=1}^{r}\alpha_k^{(v)}\varepsilon_{t-k}^2+\sum_{l=1}^{s}\beta_l^{(v)}\sigma_{t-l}^2,\quad c_{v-1}<y_{t-h}\le c_v,
    \end{aligned}
\end{equation}
where $\phi_i^{(u)}\in\eR$, $\varphi_j^{(u)}\in\eR$, $\alpha_0^{(v)}>0$, $\alpha_k^{(v)}\ge0$, $\beta_l^{(v)}\ge0$ for $k,l>0$ and $u=1,\dots,U$, $v=1,\dots,V,$ the positive integers $b,h$ are the delay parameters, the threshold parameters satisfy $-\infty=a_0<a_1<\cdots<a_U=\infty$, and $-\infty=c_0<c_1<\cdots<c_V=\infty,\{\eta_t\}$ is a sequence of i.i.d. random variables and $\eta_t$ is independent of $\cF_{t-1}$. The DTARMACH model includes many well-known models such as GARCH, TARMA \citep{brockwell1992existence}, and DTARCH \citep{li1996double}. The strict stationarity and ergodicity of the DTARMACH model were studied in \cite{ling1999probabilistic} and 
 a sufficient condition is 
\begin{flalign}\label{suff-Li}
\sum_{i=1}^{p}\max_{u}\big|\phi_i^{(u)}\big|<1\quad\mbox{and}\quad\sum_{k=1}^{r}\max_{v}\alpha_k^{(v)}+\sum_{l=1}^{s}\max_{v}\beta_l^{(v)}<1.
\end{flalign}

The DTARMACH model is a piecewise ARMA-GARCH model and can be regarded as an extension of model  \eqref{eq.ARMA-GARCH}. To incorporate model~\eqref{eq.DTARMACH} into our framework, we assume that the threshold parameters are all known, the sufficient condition (\ref{suff-Li}) holds, and $\sum_{j=1}^{q}\max_{u}|\varphi_j^{(u)}|<1$. Then $\{y_t\}$ is invertible and $\varepsilon_t$ can be written as the infinite summation of $\{y_s:s\le t\}$. For simplicity, let $p=q=r=s=1$, $d=3U+3V,$ and $\btheta=\big(\phi_0^{(1)},\phi_1^{(1)},\varphi^{(1)},\alpha_0^{(1)},\alpha_1^{(1)},\beta^{(1)},\dots,\phi_0^{(U)},\phi_1^{(U)},\varphi^{(U)},\alpha_0^{(V)},\alpha_1^{(V)},\beta^{(V)}\big)^{\T}\in\eR^{d}.$ Denote $I^{(u)}(y_{t-b})=I(a_{u-1}<y_{t-b}\le a_u)$ and $I^{(v)}(y_{t-h})=I(c_{v-1}<y_{t-h}\le c_v)$. By iterations, similar to \eqref{eq.iter_ARMA_GARCH}, it follows that
{\small
\begin{flalign*}
y_t=\sum_{k=1}^{\infty}\sum_{u_1=1}^{U}\dots\sum_{u_k=1}^{U}\prod_{l=1}^{k-1}\big\{-\varphi_1^{(u_l)}I^{(u_l)}(y_{t-b-l+1})\big\}\,\big\{\phi_0^{(u_k)}+(\phi_1^{(u_k)}+\varphi_1^{(u_k)})y_{t-k}\big\}I^{(u_k)}(y_{t-b-k+1})+\varepsilon_t,
\end{flalign*}}
where the convention $\prod_{l=1}^{0}\cdot\equiv1$ is adopted. Further, let 
{\small
\begin{flalign*}
g_t(\btheta)=\sum_{k=1}^{\infty}\sum_{u_1=1}^{U}\dots\sum_{u_k=1}^{U}\prod_{l=1}^{k-1}\{-\varphi_1^{(u_l)}I^{(u_l)}(y_{t-b-l+1})\}\{\phi_0^{(u_k)}+(\phi_1^{(u_k)}+\varphi_1^{(u_k)})y_{t-k}\}I^{(u_k)}(y_{t-b-k+1})\in \cF_{t-1},
\end{flalign*}}
and $\varepsilon_t(\btheta)=y_t-g_t(\btheta)\in \cF_{t}$. Similarly, we can get
{\small
\begin{flalign*}
\sigma_t^2(\btheta)=\sum_{k=1}^{\infty}\sum_{v_1=1}^{V}\dots\sum_{v_k=1}^{V}\prod_{l=1}^{k-1}\big\{\beta_1^{(v_l)}I^{(v_l)}(y_{t-h-l+1})\big\}\,\big\{\alpha_0^{(v_k)}+\alpha_1^{(v_k)}\varepsilon_{t-k}^2(\btheta)\big\}I^{(v_k)}(y_{t-h-k+1})\in\cF_{t-1}.
\end{flalign*}}
Thus, model~\eqref{eq.DTARMACH} has the form of model~\eqref{eq.model}. By similar arguments to Section~\ref{subsec.ARMA-GARCH}, we can verify 
Assumptions~\ref{ass.init_sigma}, \ref{ass.dot_g}, \ref{ass.Ht}, and \ref{ass.init_s_H} hold, while the others are straightforward.

\subsection{EXPAR Model}
Consider an exponential autoregressive (EXPAR) model of order $p$:
\begin{equation}
    \label{eq.EXPAR}
    y_t=\sum_{i=1}^{p}\big\{\phi_i+\varphi_i\exp(-\delta y_{t-1}^2)\big\}y_{t-i}+\eta_t,\quad t\in\eZ,
\end{equation}
where $\phi_i\in\eR$, $\varphi_i\in\eR$ for $1\leq i\leq p$, $|\phi_p|+|\varphi_p|>0$,  $\delta>0$, $\{\eta_t\}$ is a sequence of i.i.d. random variables and $\eta_t$ is independent of $\cF_{t-1}$. The EXPAR model \citep{Jones1978,haggan1981modelling} is a  nonlinear time series model partly motivated by amplitude-frequency dependency. By Theorem 2.1 in \cite{chen2018generalized}, a sufficient condition for strict stationarity and ergodicity of $\{y_t\}$ in  \eqref{eq.EXPAR} follows that all the roots of the characteristic equation: $z^p-(|\phi_1|+|\varphi_1|)z^{p-1}-\dots-(|\phi_p|+|\varphi_p|)=0$, are inside the unit circle, and $\eta_t$ has a continuous and positive density over $\eR$. 

Let $d=2p+1$, $m=p$, $\btheta=(\bphi^{\T},\bvarphi^{\T},\delta)^{\T}$ with $\bphi=(\phi_1,\dots,\phi_p)^{\T}$, $\bvarphi=(\varphi_1,\dots,\varphi_p)^{\T}$, $\bY_t=(y_t,\dots,y_{t-p+1})^{\T}$, $g(\bY_{t-1},\btheta)=\sum_{i=1}^{p}\{\phi_i+\varphi_i\exp(-\delta y_{t-1}^2)\}y_{t-i}$, and $\sigma(\bY_{t-1},\btheta)=1.$ Assumptions~\ref{ass.compact}--\ref{ass.identity}, \ref{ass.eta_2}, \ref{ass.twice}, and \ref{ass.rank} are satisfied automatically, and Assumption~\ref{ass.init_moment} is satisfied by assuming that $\eE|y_t|^{\iota}<\infty$ for some $\iota>0$. Then note that
$$
\dot{\bg}_t(\btheta)=\big(\bY_{t-1}^{\T}, ~\exp(-\delta y_{t-1}^2)\bY_{t-1}^{\T},~ -y_{t-1}^2\exp(-\delta y_{t-1}^2)\bvarphi^{\T}\bY_{t-1}\big)^{\T},
$$
and $\ddot{\bg}_t(\btheta)$ can be accordingly obtained. To guarantee Assumptions~\ref{ass.dot_g} and \ref{ass.Ht}, it only requires the conditions $\eE\Vert\dot{\bg}_t(\btheta_0)\Vert<\infty$ and $\eE\Vert\ddot{\bg}_t(\btheta_0)\Vert^2<\infty$, which can be verified by using $\eE |y_t|<\infty$ and $\eE y_t^2<\infty$, which are implied by Assumption~\ref{ass.eta} and \ref{ass.eta_2}, respectively. Thus, the LQMLE of the EXPAR model is strongly consistent and asymptotically normal.

\section{Simulation Studies}
\label{sec.simulation}
\subsection{Performance of the LQMLE}\label{subsec.sim_est}
To assess the finite-sample performance of the LQMLE,
we consider an ARMA(1,1)-GARCH(1,1) model and a DAR(1,1) model for illustrations in the following examples. 

\begin{example}\label{ex.ARMA-GARCH}
An ARMA(1,1)-GARCH(1,1) model:
\begin{equation}\label{eq.expam_AG}
        y_t=\phi_1y_{t-1}+\varepsilon_t+\varphi_1\varepsilon_{t-1}\quad{\rm with}\quad\varepsilon_t=\eta_t\sigma_t,\quad\sigma_t^2=\alpha_0+\alpha_1\varepsilon_{t-1}^2+\beta_1\sigma_{t-1}^2,\quad t\geq1,
    \end{equation}
    where $y_0=0$, $\sigma_0=0$, $\eta_t$ is generated by one of the six random variables discussed in Example~\ref{ex.psi_1}, and the true parameter $\btheta_0=(\phi_1,\varphi_1,\alpha_0,\alpha_1,\beta_1)^{\T}$ $=(0.3,0.2,0.2,0.1,0.3)^{\T}$ and $(0.2,0.3,0.3,0.1,0.2)^{\T}$ in Scenario I and Scenario II, respectively. 
\end{example}

\begin{example}\label{ex.DAR}
A DAR(1,1) model:
\begin{equation}\label{eq.expam_DAR}
y_t=\phi_0+\phi_1y_{t-1}+\eta_t(\alpha_0+\alpha_1y_{t-1}^2)^{1/2},\quad t\geq1,
\end{equation}
where $y_0=0,\eta_t$ is generated by one of the six random variables discussed in Example~\ref{ex.psi_1}, and the true parameter $\btheta_0=(\phi_0,\phi_1,\alpha_0,\alpha_1)^{\T}=(1.0,0.5,0.3,0.5)^{\T}$ and $(0.5,0.2,1.0,0.3)^{\T}$ in Scenario I and Scenario II, respectively. 
\end{example}

In each simulation scenario, we use
the length of observations $n=100$, 200, and 400 with $1000$ replications. 
The absolute estimation bias and standard deviation results for each simulation experiment of models~\eqref{eq.expam_AG} and \eqref{eq.expam_DAR} are reported in Tables \ref{tab.est_AG} and \ref{tab.est_DAR}, respectively. From the tables, we can find that (i)
with a larger $n$, both the absolute estimation bias and standard deviation of each parameter show decreasing trends; 
(ii) even though it is motivated by the log-likelihood function of a standard logistic distribution when the model innovations come from other distributions, our LQMLE is still consistent, which implies that the LQMLE is robust to distributional misspecifications.

\begin{table}[H]
	\caption{The absolute estimation bias and standard deviation results for model~\eqref{eq.expam_AG}.}
        \scriptsize
        \vspace{-1em}
	\label{tab.est_AG}
	\begin{center}
		\begin{tabular}{lcccccccccc}
			\toprule
            & \multicolumn{5}{l}{Scenario I} &  \multicolumn{5}{l}{Scenario II}\\
			$n$ & $\phi_1=0.3$ & $\varphi_1=0.2$ & $\alpha_0=0.2$ & $\alpha_1=0.1$ & $\beta_1=0.3$ & $\phi_1=0.2$ & $\varphi_1=0.3$ & $\alpha_0=0.3$ & $\alpha_1=0.1$ & $\beta_1=0.2$ \\
			\hline
			& \multicolumn{10}{l}{$\eta_t\sim{\rm Logistic}(0,1)$} \\
                100   & 0.0215  & 0.0213  & 0.0439  & 0.0051  & 0.0710  & 0.0050  & 0.0007  & 0.0415  & 0.0040  & 0.0636  \\
          & (0.232) & (0.236) & (0.159) & (0.080) & (0.374) & (0.243) & (0.253) & (0.197) & (0.078) & (0.380) \\
    200   & 0.0072  & 0.0062  & 0.0306  & 0.0036  & 0.0527  & 0.0043  & 0.0041  & 0.0293  & 0.0016  & 0.0506  \\
          & (0.163) & (0.169) & (0.124) & (0.049) & (0.289) & (0.160) & (0.163) & (0.153) & (0.050) & (0.292) \\
    400   & 0.0033  & 0.0033  & 0.0148  & 0.0004  & 0.0293  & 0.0015  & 0.0030  & 0.0143  & 0.0015  & 0.0201  \\
          & (0.106) & (0.109) & (0.079) & (0.033) & (0.193) & (0.105) & (0.105) & (0.098) & (0.034) & (0.189) \\
          \hline

          & \multicolumn{10}{l}{$\eta_t\sim\cN(0,1.75^2)$} \\
          100   & 0.0132  & 0.0092  & 0.0381  & 0.0079  & 0.0641  & 0.0001  & 0.0026  & 0.0421  & 0.0048  & 0.0618  \\
          & (0.246) & (0.258) & (0.154) & (0.071) & (0.379) & (0.254) & (0.263) & (0.197) & (0.076) & (0.390) \\
    200   & 0.0161  & 0.0155  & 0.0216  & 0.0005  & 0.0417  & 0.0041  & 0.0033  & 0.0303  & 0.0059  & 0.0350  \\
          & (0.166) & (0.165) & (0.106) & (0.046) & (0.265) & (0.165) & (0.159) & (0.159) & (0.047) & (0.304) \\
    400   & 0.0003  & 0.0017  & 0.0164  & 0.0020  & 0.0274  & 0.0051  & 0.0028  & 0.0129  & 0.0004  & 0.0194  \\
          & (0.107) & (0.114) & (0.074) & (0.030) & (0.180) & (0.112) & (0.109) & (0.109) & (0.032) & (0.211) \\
          \hline

          & \multicolumn{10}{l}{$\eta_t\sim U(-2.85,2.85)$} \\
    100   & 0.0231  & 0.0141  & 0.0431  & 0.0203  & 0.0602  & 0.0007  & 0.0027  & 0.0514  & 0.0200  & 0.0582  \\
          & (0.269) & (0.276) & (0.151) & (0.072) & (0.418) & (0.264) & (0.274) & (0.205) & (0.072) & (0.439) \\
    200   & 0.0038  & 0.0055  & 0.0330  & 0.0091  & 0.0600  & 0.0078  & 0.0036  & 0.0339  & 0.0096  & 0.0514  \\
          & (0.175) & (0.183) & (0.121) & (0.041) & (0.307) & (0.179) & (0.181) & (0.170) & (0.040) & (0.341) \\
    400   & 0.0009  & 0.0023  & 0.0132  & 0.0033  & 0.0265  & 0.0028  & 0.0036  & 0.0214  & 0.0027  & 0.0354  \\
          & (0.116) & (0.117) & (0.074) & (0.026) & (0.183) & (0.121) & (0.120) & (0.113) & (0.026) & (0.220) \\
          \hline

          & \multicolumn{10}{l}{$\eta_t\sim 1.25 t_3$} \\
    100   & 0.0159  & 0.0127  & 0.0410  & 0.0080  & 0.0591  & 0.0032  & 0.0033  & 0.0270  & 0.0072  & 0.0464  \\
          & (0.225) & (0.239) & (0.160) & (0.108) & (0.342) & (0.223) & (0.227) & (0.187) & (0.105) & (0.350) \\
    200   & 0.0153  & 0.0140  & 0.0258  & 0.0087  & 0.0505  & 0.0007  & 0.0068  & 0.0266  & 0.0065  & 0.0344  \\
          & (0.163) & (0.166) & (0.117) & (0.073) & (0.262) & (0.156) & (0.157) & (0.141) & (0.071) & (0.263) \\
    400   & 0.0049  & 0.0053  & 0.0117  & 0.0041  & 0.0218  & 0.0021  & 0.0026  & 0.0089  & 0.0031  & 0.0101  \\
          & (0.107) & (0.112) & (0.076) & (0.049) & (0.182) & (0.103) & (0.101) & (0.101) & (0.047) & (0.187) \\
          \hline

          & \multicolumn{10}{l}{$\eta_t\sim 0.96 t_2$} \\
    100   & 0.0146  & 0.0114  & 0.0336  & 0.0395  & 0.0580  & 0.0046  & 0.0126  & 0.0198  & 0.0223  & 0.0085  \\
          & (0.234) & (0.241) & (0.200) & (0.214) & (0.333) & (0.223) & (0.211) & (0.310) & (0.193) & (0.358) \\
    200   & 0.0094  & 0.0107  & 0.0164  & 0.0183  & 0.0395  & 0.0035  & 0.0040  & 0.0076  & 0.0324  & 0.0239  \\
          & (0.154) & (0.160) & (0.126) & (0.128) & (0.266) & (0.161) & (0.157) & (0.157) & (0.152) & (0.256) \\
    400   & 0.0004  & 0.0025  & 0.0149  & 0.0187  & 0.0388  & 0.0035  & 0.0044  & 0.0043  & 0.0213  & 0.0186  \\
          & (0.113) & (0.124) & (0.103) & (0.118) & (0.193) & (0.110) & (0.108) & (0.121) & (0.119) & (0.193) \\
          \hline

          & \multicolumn{10}{l}{$\eta_t\sim S(1.69,0,1,0)$} \\
    100   & 0.0233  & 0.0130  & 0.0342  & 0.0211  & 0.0653  & 0.0034  & 0.0032  & 0.0242  & 0.0224  & 0.0523  \\
          & (0.249) & (0.257) & (0.199) & (0.185) & (0.348) & (0.245) & (0.247) & (0.217) & (0.191) & (0.352) \\
    200   & 0.0028  & 0.0058  & 0.0195  & 0.0192  & 0.0494  & 0.0033  & 0.0072  & 0.0047  & 0.0164  & 0.0225  \\
          & (0.178) & (0.184) & (0.157) & (0.150) & (0.260) & (0.176) & (0.174) & (0.170) & (0.149) & (0.267) \\
    400   & 0.0046  & 0.0042  & 0.0063  & 0.0133  & 0.0312  & 0.0008  & 0.0012  & 0.0057  & 0.0059  & 0.0231  \\
          & (0.119) & (0.126) & (0.083) & (0.118) & (0.184) & (0.113) & (0.116) & (0.120) & (0.086) & (0.200) \\
	   
			\bottomrule
		\end{tabular}
	\end{center}
    \vspace{-2em}
\end{table}

\begin{table}[H]
	\caption{The absolute estimation bias and standard deviation results for model~\eqref{eq.expam_DAR}.}
        \scriptsize
        \vspace{-1em}
	\label{tab.est_DAR}
	\begin{center}
		\begin{tabular}{lcccccccc}
			\toprule
            & \multicolumn{4}{l}{Scenario I} &  \multicolumn{4}{l}{Scenario II}\\
			$n$ & $\phi_0=1.0$ & $\phi_1=0.5$ & $\alpha_0=0.3$ & $\alpha_1=0.5$ & $\phi_0=0.5$ & $\phi_1=0.2$ & $\alpha_0=1.0$ & $\alpha_1=0.3$ \\
			\hline
			& \multicolumn{8}{l}{$\eta_t\sim{\rm Logistic}(0,1)$} \\
                100   & 0.0217  & 0.0041  & 0.0055  & 0.0080  & 0.0056  & 0.0057  & 0.0269  & 0.0193  \\
          & (0.220) & (0.146) & (0.166) & (0.108) & (0.248) & (0.137) & (0.329) & (0.096) \\
    200   & 0.0146  & 0.0021  & 0.0058  & 0.0055  & 0.0005  & 0.0034  & 0.0294  & 0.0121  \\
          & (0.159) & (0.098) & (0.115) & (0.073) & (0.173) & (0.097) & (0.217) & (0.071) \\
    400   & 0.0074  & 0.0039  & 0.0014  & 0.0009  & 0.0058  & 0.0028  & 0.0044  & 0.0041  \\
          & (0.105) & (0.069) & (0.076) & (0.053) & (0.123) & (0.067) & (0.153) & (0.045) \\
          \hline

          & \multicolumn{8}{l}{$\eta_t\sim\cN(0,1.75^2)$} \\
          100   & 0.0339  & 0.0061  & 0.0027  & 0.0071  & 0.0150  & 0.0152  & 0.0283  & 0.0160  \\
          & (0.234) & (0.149) & (0.144) & (0.092) & (0.255) & (0.143) & (0.295) & (0.086) \\
    200   & 0.0152  & 0.0131  & 0.0011  & 0.0067  & 0.0037  & 0.0009  & 0.0128  & 0.0083  \\
          & (0.161) & (0.107) & (0.099) & (0.066) & (0.181) & (0.099) & (0.201) & (0.058) \\
    400   & 0.0105  & 0.0032  & 0.0007  & 0.0029  & 0.0019  & 0.0027  & 0.0099  & 0.0051  \\
          & (0.110) & (0.071) & (0.065) & (0.046) & (0.123) & (0.069) & (0.137) & (0.041) \\
          \hline

          & \multicolumn{8}{l}{$\eta_t\sim U(-2.85,2.85)$} \\
    100   & 0.0484  & 0.0208  & 0.0198  & 0.0141  & 0.0033  & 0.0148  & 0.0017  & 0.0169  \\
          & (0.254) & (0.161) & (0.095) & (0.064) & (0.297) & (0.151) & (0.218) & (0.058) \\
    200   & 0.0209  & 0.0118  & 0.0060  & 0.0064  & 0.0009  & 0.0061  & 0.0078  & 0.0072  \\
          & (0.175) & (0.110) & (0.068) & (0.046) & (0.213) & (0.102) & (0.150) & (0.042) \\
    400   & 0.0120  & 0.0093  & 0.0035  & 0.0033  & 0.0009  & 0.0033  & 0.0067  & 0.0048  \\
          & (0.123) & (0.079) & (0.048) & (0.031) & (0.141) & (0.072) & (0.103) & (0.029) \\
          \hline

          & \multicolumn{8}{l}{$\eta_t\sim 1.25 t_3$} \\
    100   & 0.0140  & 0.0000  & 0.0201  & 0.0060  & 0.0183  & 0.0170  & 0.0498  & 0.0105  \\
          & (0.209) & (0.134) & (0.277) & (0.177) & (0.233) & (0.133) & (0.463) & (0.151) \\
    200   & 0.0169  & 0.0070  & 0.0139  & 0.0022  & 0.0078  & 0.0083  & 0.0313  & 0.0068  \\
          & (0.139) & (0.097) & (0.179) & (0.110) & (0.155) & (0.094) & (0.292) & (0.106) \\
    400   & 0.0054  & 0.0010  & 0.0094  & 0.0006  & 0.0013  & 0.0043  & 0.0298  & 0.0019  \\
          & (0.097) & (0.068) & (0.114) & (0.079) & (0.113) & (0.067) & (0.219) & (0.074) \\
          \hline

          & \multicolumn{8}{l}{$\eta_t\sim 0.96 t_2$} \\
    100   & 0.0290  & 0.0113  & 0.1099  & 0.0331  & 0.0167  & 0.0186  & 0.0740  & 0.0010  \\
          & (0.191) & (0.129) & (0.830) & (0.834) & (0.196) & (0.127) & (0.747) & (0.291) \\
    200   & 0.0112  & 0.0028  & 0.0337  & 0.0138  & 0.0149  & 0.0097  & 0.0774  & 0.0065  \\
          & (0.129) & (0.086) & (0.332) & (0.426) & (0.134) & (0.086) & (0.673) & (0.298) \\
    400   & 0.0046  & 0.0037  & 0.0276  & 0.0120  & 0.0020  & 0.0051  & 0.0495  & 0.0019  \\
          & (0.090) & (0.065) & (0.254) & (0.200) & (0.094) & (0.064) & (0.452) & (0.174) \\
          \hline

          & \multicolumn{8}{l}{$\eta_t\sim S(1.69,0,1,0)$} \\
    100   & 0.0271  & 0.0080  & 0.0739  & 0.0132  & 0.0160  & 0.0132  & 0.1194  & 0.0043  \\
          & (0.199) & (0.137) & (0.763) & (0.266) & (0.217) & (0.132) & (1.012) & (0.294) \\
    200   & 0.0188  & 0.0008  & 0.0311  & 0.0185  & 0.0028  & 0.0079  & 0.0570  & 0.0023  \\
          & (0.135) & (0.094) & (0.576) & (0.439) & (0.158) & (0.092) & (0.774) & (0.224) \\
    400   & 0.0041  & 0.0019  & 0.0142  & 0.0068  & 0.0016  & 0.0042  & 0.0054  & 0.0018  \\
          & (0.097) & (0.066) & (0.226) & (0.158) & (0.105) & (0.063) & (0.382) & (0.194) \\
	   
			\bottomrule
		\end{tabular}
	\end{center}
    \vspace{-2em}
\end{table}

\begin{figure}[H]
\begin{center}
\includegraphics[width=0.63\linewidth]{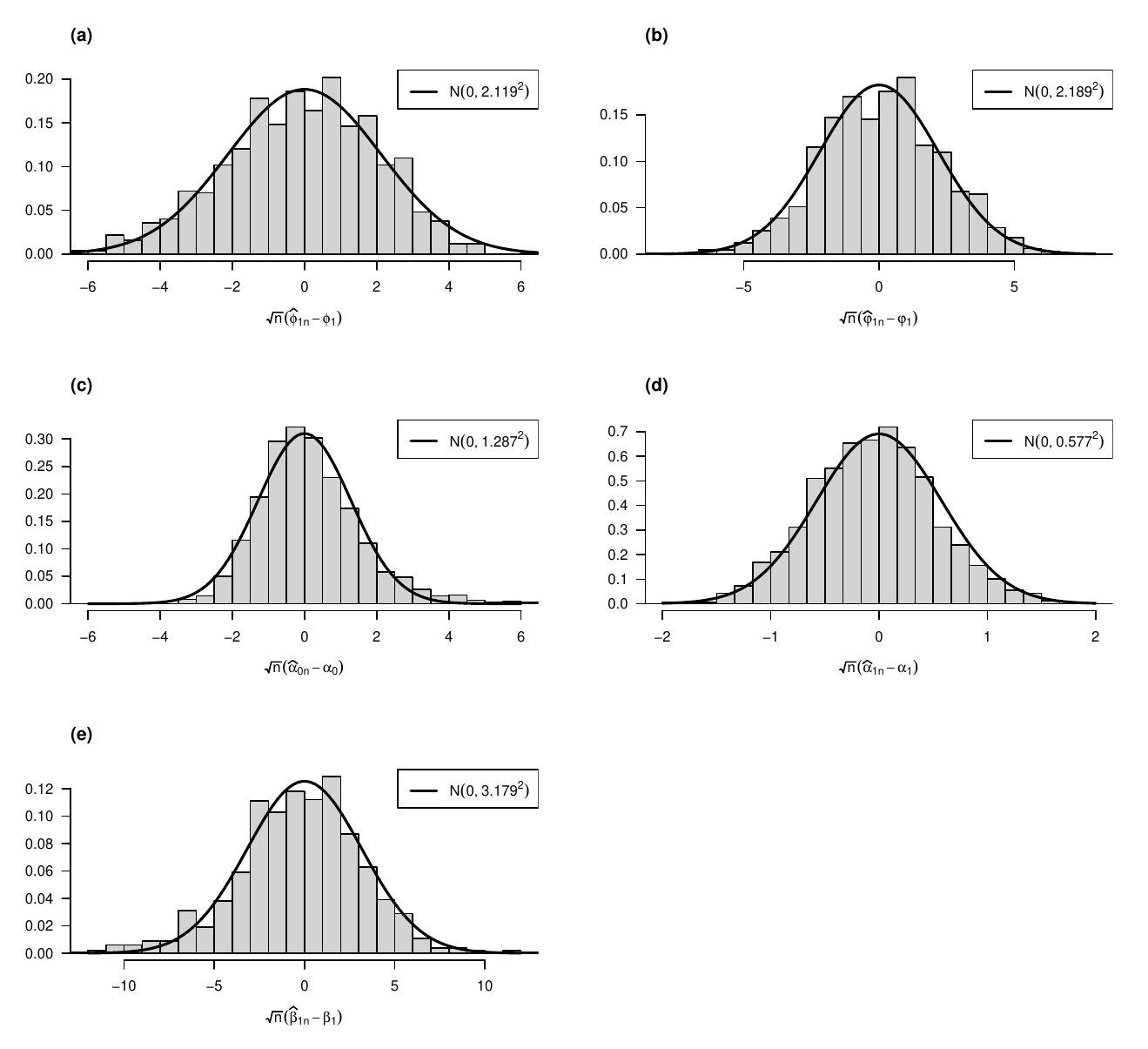}
\end{center}
\vspace{-1em}
\caption{The histograms of $\sqrt{n}(\widehat{\btheta}_n-\btheta_0)$ for model~\eqref{eq.expam_AG} with $n=400$, $\eta_t\sim\cN(0,1.75^2)$, and the true parameter $\btheta_0=(0.3,0.2,0.2,0.1,0.3)^{\T}$.}
\vspace{-1em}
\label{fig.hist_ARMA-GARCH_N}
\end{figure}

\begin{figure}[H]
\begin{center}
\includegraphics[width=0.63\linewidth]{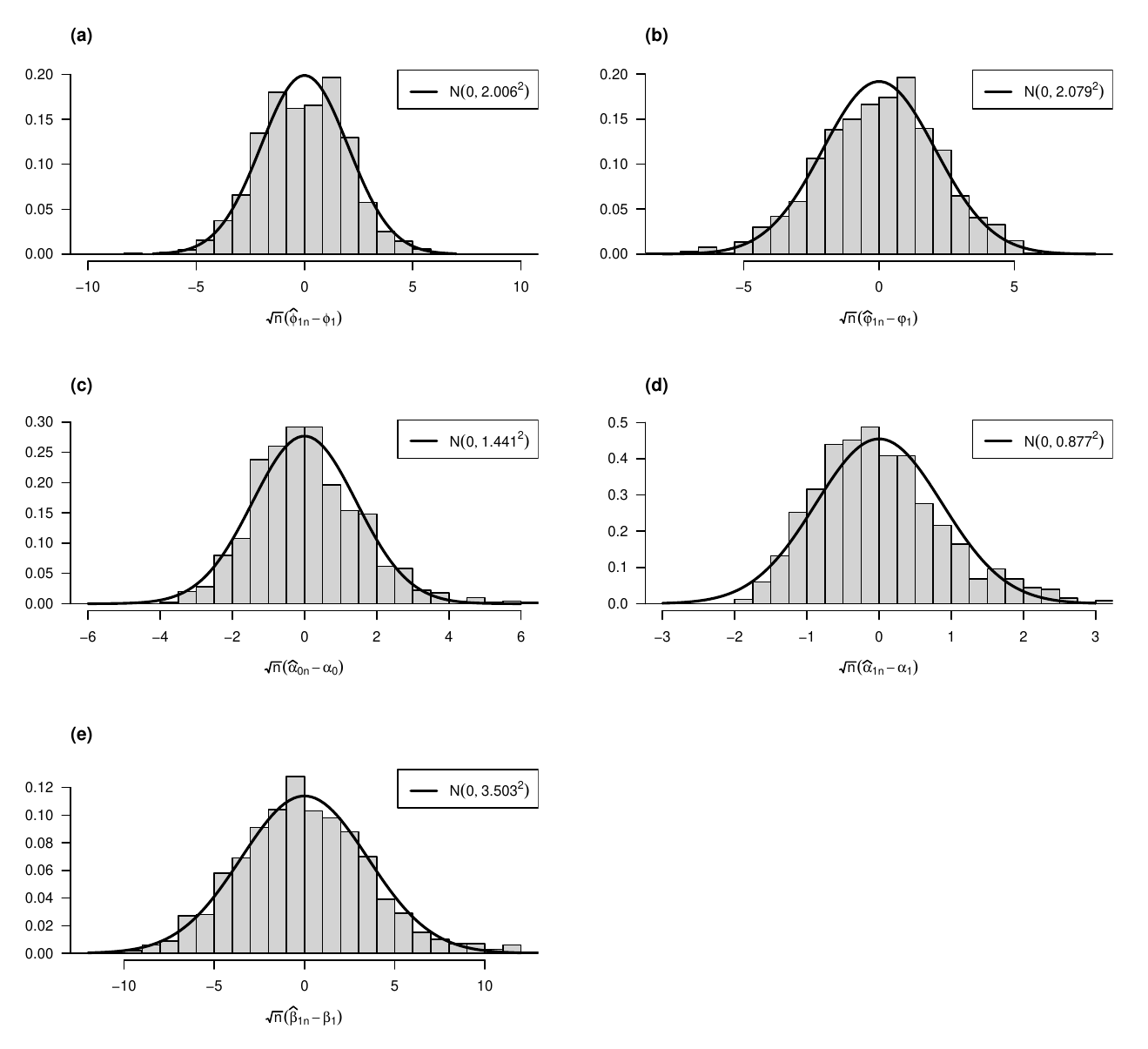}
\end{center}
\vspace{-1em}
\caption{The histograms of $\sqrt{n}(\widehat{\btheta}_n-\btheta_0)$ for model~\eqref{eq.expam_AG} with $n=400$, $\eta_t\sim1.25 t_3$, and the true parameter $\btheta_0=(0.3,0.2,0.2,0.1,0.3)^{\T}$.}
\vspace{-1em}
\label{fig.hist_ARMA-GARCH_t}
\end{figure}

\begin{figure}[H]
\begin{center}
\includegraphics[width=0.63\linewidth]{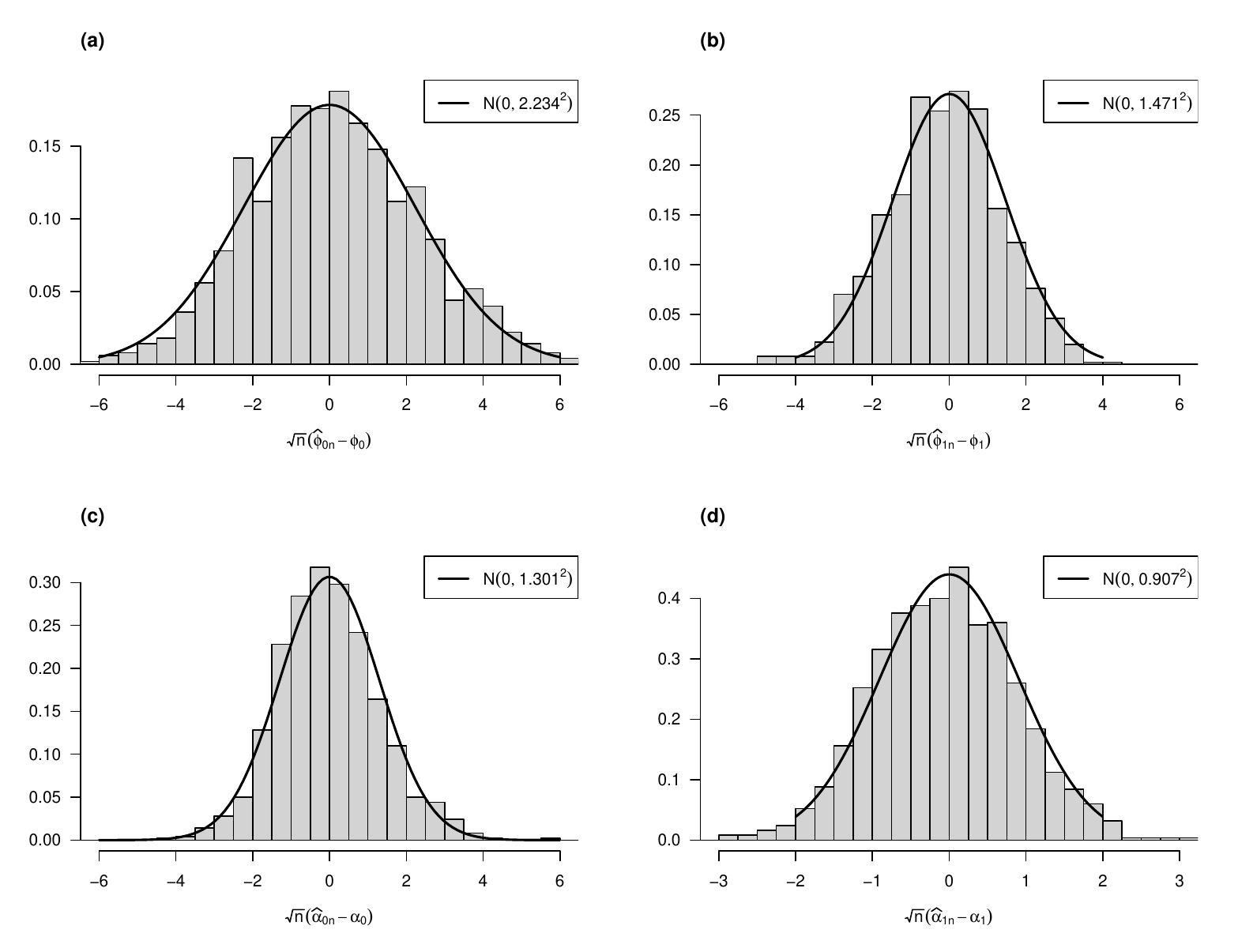}
\end{center}
\vspace{-1em}
\caption{The histograms of $\sqrt{n}(\widehat{\btheta}_n-\btheta_0)$ for model~\eqref{eq.expam_DAR} with $n=400$, $\eta_t\sim\cN(0,1.75^2)$, and the true parameter $\btheta_0=(1.0,0.5,0.3,0.5)^{\T}$.}
\vspace{-1em}
\label{fig.hist_DAR_N}
\end{figure}

\begin{figure}[H]
\begin{center}
\includegraphics[width=0.63\linewidth]{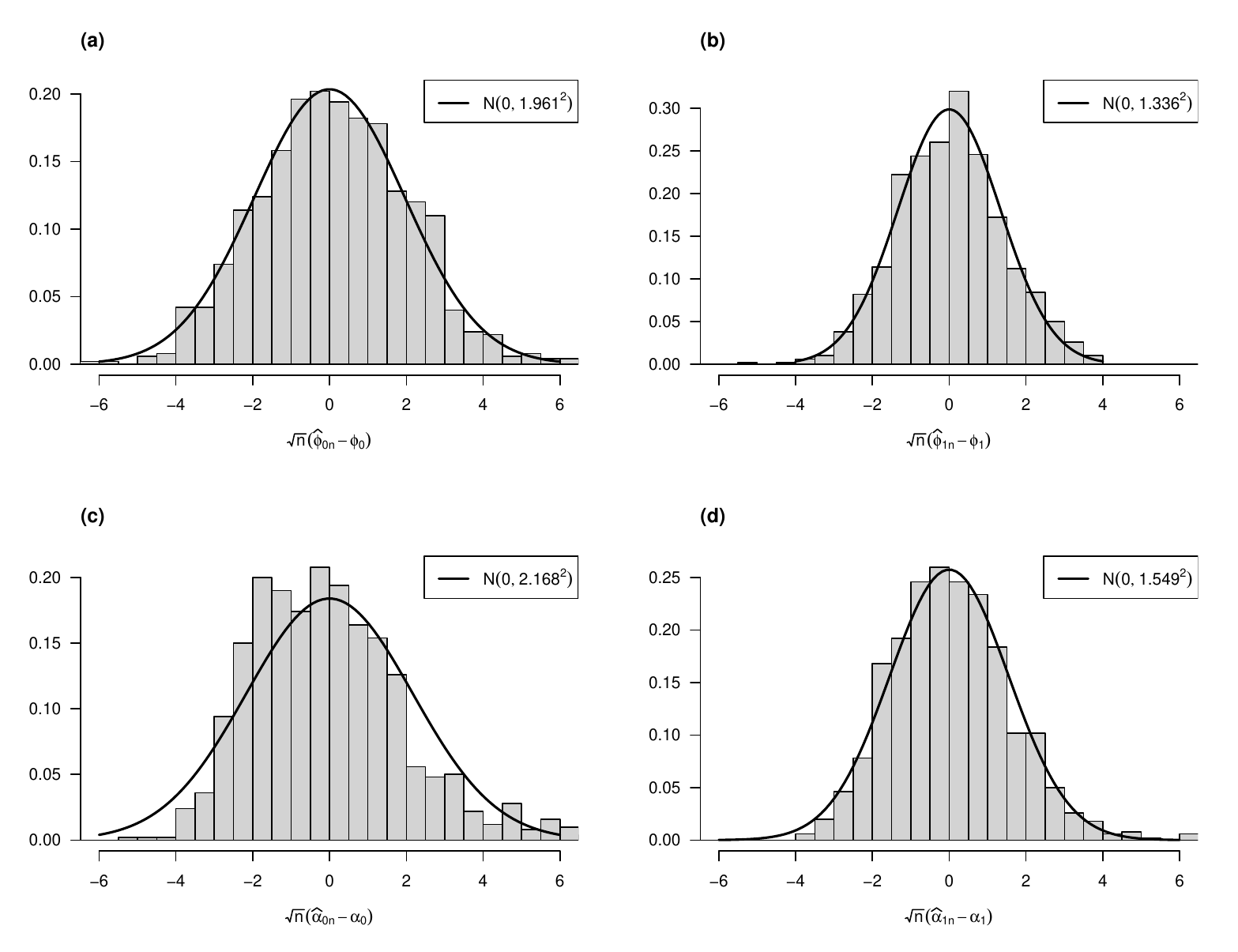}
\end{center}
\vspace{-1em}
\caption{The histograms of $\sqrt{n}(\widehat{\btheta}_n-\btheta_0)$ for model~\eqref{eq.expam_DAR} with $n=400$, $\eta_t\sim 1.25 t_3$ and the true parameter $\btheta_0=(1.0,0.5,0.3,0.5)^{\T}$.}
\vspace{-1em}
\label{fig.hist_DAR_t}
\end{figure}

Figures~\ref{fig.hist_ARMA-GARCH_N}--\ref{fig.hist_DAR_t} plot the histograms of $\sqrt{n}(\widehat{\btheta}_n-\btheta_0)$ for models~\eqref{eq.expam_AG} and \eqref{eq.expam_DAR} with $n=400$, the innovations $\{\eta_t\}$ being generated from i.i.d. $\cN(0,1.75^2)$ and $1.25 t_3$, and the true parameter $(\phi_1,\varphi_1,\alpha_0,\alpha_1,\beta_1)=(0.3,0.2,0.2,0.1,0.3)$ and $(\phi_0,\phi_1,\alpha_0,\alpha_1)=(1.0,0.5,0.3,0.5)$, respectively. The asymptotic standard deviations are calculated from the
asymptotic covariance matrix in Theorems~\ref{thm.clt} through Monte Carlo simulation. It can be seen that the empirical densities of each estimator closely align with the asymptotic normal distributions, despite the absence of finite fourth moments of the innovations, which are generated from a Student's $t_{3}$-distribution. These findings strongly support Theorems~\ref{thm.clt} and demonstrate the robustness of the LQMLE compared to the GQMLE.

Finally, we compare the finite-sample performance of our LQMLE and the GQMLE. As discussed in Remark~\ref{rmk.ass_consistency}, the identifiability conditions for the LQMLE and GQMLE are different so that it is infeasible to fairly compare the finite-sample performance of the LQMLE of parameters in volatility function directly. Thus, we first focus on the estimators of parameters in mean functions, i.e., $(\phi_1,\varphi_1)$ in ARMA-GARCH model~\eqref{eq.expam_AG}, and $(\phi_0,\phi_1)$ in DAR model~\eqref{eq.expam_DAR}. In each simulation scenario, we use $n=100,200,400$ with $1000$ replications. The absolute estimation bias and standard deviation results for two estimators are reported in Table~\ref{tab.compare}.  
\begin{table}[!htbp]
	\caption{The absolute estimation bias and standard deviation results of the LQMLE and GQMLE for Scenario I of model~\eqref{eq.expam_AG} and Scenario I of model~\eqref{eq.expam_DAR}.}
        \scriptsize
	\label{tab.compare}
	\begin{center}
		\begin{tabular}{lcccccccc}
			\toprule
            & \multicolumn{4}{l}{Scenario I of model~\eqref{eq.expam_AG}} &  \multicolumn{4}{l}{Scenario I of model~\eqref{eq.expam_DAR}}\\
            & \multicolumn{2}{l}{$\phi_1=0.3$} &  \multicolumn{2}{l}{$\varphi_1=0.2$} &  \multicolumn{2}{l}{$\phi_0=1.0$} &  \multicolumn{2}{l}{$\phi_1=0.5$}\\
			$n$ & LQMLE & GQMLE & LQMLE & GQMLE & LQMLE & GQMLE & LQMLE & GQMLE \\
			\hline
			& \multicolumn{8}{l}{$\eta_t\sim{\rm Logistic}(0,1)$} \\
                100   & 0.0023  & 0.0033  & 0.0005  & 0.0043  & 0.0368  & 0.0311  & 0.0027  & 0.0007  \\
          & (0.232) & (0.242) & (0.248) & (0.261) & (0.231) & (0.241) & (0.145) & (0.151) \\
    200   & 0.0103  & 0.0087  & 0.0087  & 0.0063  & 0.0172  & 0.0168  & 0.0034  & 0.0023  \\
          & (0.156) & (0.167) & (0.167) & (0.178) & (0.154) & (0.162) & (0.099) & (0.103) \\
    400   & 0.0002  & 0.0007  & 0.0031  & 0.0039  & 0.0067  & 0.0050  & 0.0008  & 0.0003  \\
          & (0.107) & (0.113) & (0.113) & (0.120) & (0.113) & (0.117) & (0.072) & (0.075) \\
          \hline

          & \multicolumn{8}{l}{$\eta_t\sim\cN(0,1.75^2)$} \\
          100   & 0.0054  & 0.0027  & 0.0016  & 0.0043  & 0.0274  & 0.0232  & 0.0187  & 0.0183  \\
          & (0.241) & (0.234) & (0.257) & (0.247) & (0.237) & (0.230) & (0.149) & (0.145) \\
    200   & 0.0030  & 0.0013  & 0.0000  & 0.0025  & 0.0179  & 0.0188  & 0.0043  & 0.0042  \\
          & (0.165) & (0.160) & (0.171) & (0.166) & (0.157) & (0.154) & (0.104) & (0.102) \\
    400   & 0.0073  & 0.0057  & 0.0057  & 0.0030  & 0.0069  & 0.0064  & 0.0032  & 0.0022  \\
          & (0.109) & (0.105) & (0.114) & (0.110) & (0.113) & (0.111) & (0.076) & (0.074) \\
          \hline

          & \multicolumn{8}{l}{$\eta_t\sim U(-2.85,2.85)$} \\
    100   & 0.0239  & 0.0200  & 0.0185  & 0.0153  & 0.0416  & 0.0351  & 0.0121  & 0.0097  \\
          & (0.274) & (0.234) & (0.277) & (0.239) & (0.258) & (0.222) & (0.151) & (0.129) \\
    200   & 0.0017  & 0.0009  & 0.0062  & 0.0041  & 0.0215  & 0.0176  & 0.0036  & 0.0031  \\
          & (0.174) & (0.143) & (0.186) & (0.153) & (0.176) & (0.149) & (0.113) & (0.096) \\
    400   & 0.0081  & 0.0067  & 0.0043  & 0.0035  & 0.0133  & 0.0104  & 0.0041  & 0.0037  \\
          & (0.119) & (0.099) & (0.120) & (0.101) & (0.127) & (0.106) & (0.076) & (0.064) \\
          \hline

          & \multicolumn{8}{l}{$\eta_t\sim 1.25 t_3$} \\
    100   & 0.0187  & 0.0272  & 0.0168  & 0.0031  & 0.0363  & 0.0353  & 0.0065  & 0.0058  \\
          & (0.232) & (0.332) & (0.241) & (0.336) & (0.215) & (0.272) & (0.135) & (0.183) \\
    200   & 0.0117  & 0.0076  & 0.0087  & 0.0074  & 0.0115  & 0.0100  & 0.0006  & 0.0003  \\
          & (0.154) & (0.237) & (0.168) & (0.266) & (0.149) & (0.200) & (0.093) & (0.125) \\
    400   & 0.0015  & 0.0060  & 0.0007  & 0.0127  & 0.0023  & 0.0021  & 0.0035  & 0.0017  \\
          & (0.103) & (0.190) & (0.108) & (0.213) & (0.101) & (0.131) & (0.069) & (0.092) \\
          \hline

          & \multicolumn{8}{l}{$\eta_t\sim 0.96 t_2$} \\
    100   & 0.0105  & 0.0573  & 0.0018  & 0.0055  & 0.0188  & 0.0563  & 0.0104  & 0.0172  \\
          & (0.220) & (0.404) & (0.234) & (0.375) & (0.244) & (2.323) & (0.127) & (0.580) \\
    200   & 0.0018  & 0.0224  & 0.0028  & 0.0506  & 0.0153  & 0.0174  & 0.0067  & 0.0219  \\
          & (0.168) & (0.440) & (0.175) & (0.392) & (0.128) & (0.282) & (0.090) & (0.239) \\
    400   & 0.0037  & 0.0520  & 0.0011  & 0.0267  & 0.0068  & 0.0199  & 0.0012  & 0.0032  \\
          & (0.113) & (0.386) & (0.119) & (0.366) & (0.090) & (0.453) & (0.062) & (0.153) \\
          \hline

          & \multicolumn{8}{l}{$\eta_t\sim S(1.69,0,1,0)$} \\
    100   & 0.0207  & 0.0673  & 0.0183  & 0.0100  & 0.0197  & 0.0277  & 0.0084  & 0.0107  \\
          & (0.242) & (0.651) & (0.256) & (0.372) & (0.202) & (0.430) & (0.138) & (0.262) \\
    200   & 0.0065  & 0.0332  & 0.0045  & 0.0459  & 0.0120  & 0.0121  & 0.0070  & 0.0173  \\
          & (0.169) & (0.474) & (0.174) & (0.350) & (0.134) & (0.430) & (0.096) & (0.494) \\
    400   & 0.0032  & 0.0230  & 0.0000  & 0.0323  & 0.0094  & 0.0233  & 0.0048  & 0.0048  \\
          & (0.120) & (0.335) & (0.118) & (0.320) & (0.094) & (0.290) & (0.067) & (0.178) \\
			\bottomrule
		\end{tabular}
	\end{center}
    \vspace{-2em}
\end{table}

From the table, we can see that (i) when $\eta_t$ follows the logistic distribution, the LQMLE is indeed the MLE and its outperformance is clear; (ii) when $\eta_t$ follows the normal distribution, the GQMLE reduces to the MLE and outperforms the LQMLE, which is unquestionable; (iii) when $\eta_t$ follows the uniform distribution, the GQMLE outperforms the LQMLE since the uniform distribution is lighted-tailed; (iv) when $\eta_t$ follows the Student's $t_\nu$ and stable distributions, the LQMLE achieves significantly better performance than the GQMLE, which shows that the LQMLE is robust to heavy-tailed innovations. Moreover, the heavier the tail of the innovation, the better the performance of the LQMLE.

To fairly compare the finite-sample performance of our LQMLE and the GQMLE of the parameters in the volatility function, a rescaling technique \citep{li2018zd} can be used to address the issues caused by different identifiability conditions. The related numerical results are presented in Section~{\color{blue}S.4.1} of the supplementary material, where we have similar findings.

\subsection{Performance of the Test Statistics}\label{subsec.sim_test}
To examine the finite-sample performance of two tests proposed in Section~\ref{subsec.inference}, we consider the null hypothesis $H_0:\btheta_0=\btheta_{H_0}=(\phi_{1,H_0},\varphi_{1,H_0},\alpha_{0,H_0},\alpha_{1,H_0},\beta_{1,H_0})^{\T}=(0.3,0.2,0.2,0.1,0.3)^{\T}$ and $\bR=(1,1,2,3,1)$ for model~\eqref{eq.expam_AG} , and $H_0:\btheta_0=\btheta_{H_0}=(\phi_{0,H_0},\phi_{1,H_0},\alpha_{0,H_0},\alpha_{1,H_0})^{\T}=(1.0,0.5,0.3,0.5)^{\T}$ and $\bR=(1,1,1,1)$ for model~\eqref{eq.expam_DAR}. As discussed in Remark~\ref{rmk.ass_nor}, the asymptotic normality of the estimator requires the innovations with finite variance, thus $\eta_t$ is generated by one of the first four random variables discussed in Example~\ref{ex.psi_1}. The nominal significance level is $\alpha=5\%.$ To present the empirical powers, we consider three alternatives $H_1:\btheta_0=1.1\btheta_{H_0}$, $H_1:\btheta_0=1.3\btheta_{H_0}$, and $H_1:\btheta_0=1.5\btheta_{H_0}$ for both two models, respectively. In each simulation scenario, we use $n=100,200,400$ with $1000$ replications. The empirical sizes and powers of the hypothesis tests for models~\eqref{eq.expam_AG} and \eqref{eq.expam_DAR} are reported in Tables~\ref{tab.infer_AG} and \ref{tab.infer_DAR}, respectively. 
\begin{table}[!htbp]
	\caption{The empirical sizes and powers of the hypothesis tests for model~\eqref{eq.expam_AG} at $\alpha=5\%$.}
        \scriptsize
        \vspace{-1em}
	\label{tab.infer_AG}
	\begin{center}
		\begin{tabular}{lcccccccc}
			\toprule
            & \multicolumn{2}{l}{$\btheta_0=\btheta_{H_0}$} &  \multicolumn{2}{l}{$\btheta_0=1.1\btheta_{H_0}$} & \multicolumn{2}{l}{$\btheta_0=1.3\btheta_{H_0}$} & \multicolumn{2}{l}{$\btheta_0=1.5\btheta_{H_0}$} \\
			$n$ & Wald & LM & Wald & LM & Wald & LM & Wald & LM \\
			\hline
			& \multicolumn{8}{l}{$\eta_t\sim{\rm Logistic}(0,1)$} \\
                    100   & 0.041  & 0.031  & 0.069  & 0.069  & 0.420  & 0.460  & 0.741  & 0.821  \\
    200   & 0.022  & 0.018  & 0.150  & 0.155  & 0.771  & 0.837  & 0.933  & 0.976  \\
    400   & 0.042  & 0.041  & 0.317  & 0.331  & 0.975  & 0.993  & 0.992  & 0.998  \\
          \hline

          & \multicolumn{8}{l}{$\eta_t\sim\cN(0,1.75^2)$} \\
          100   & 0.040  & 0.024  & 0.072  & 0.050  & 0.491  & 0.463  & 0.790  & 0.865  \\
    200   & 0.034  & 0.031  & 0.172  & 0.150  & 0.788  & 0.829  & 0.952  & 0.982  \\
    400   & 0.049  & 0.039  & 0.368  & 0.355  & 0.975  & 0.993  & 0.998  & 1.000  \\
          \hline

          & \multicolumn{8}{l}{$\eta_t\sim U(-2.85,2.85)$} \\
    100   & 0.033  & 0.038  & 0.147  & 0.050  & 0.595  & 0.483  & 0.833  & 0.864  \\
    200   & 0.048  & 0.043  & 0.237  & 0.132  & 0.854  & 0.813  & 0.964  & 0.979  \\
    400   & 0.028  & 0.040  & 0.491  & 0.392  & 0.976  & 0.986  & 0.999  & 0.999  \\
            \hline

          & \multicolumn{8}{l}{$\eta_t\sim 1.25 t_3$} \\
    100   & 0.037  & 0.033  & 0.033  & 0.070  & 0.272  & 0.419  & 0.606  & 0.788  \\
    200   & 0.039  & 0.032  & 0.063  & 0.120  & 0.603  & 0.765  & 0.859  & 0.962  \\
    400   & 0.039  & 0.032  & 0.164  & 0.256  & 0.914  & 0.978  & 0.982  & 0.996  \\

			\bottomrule
		\end{tabular}
	\end{center}
    \vspace{-2em}
\end{table}
\begin{table}[H]
	\caption{The empirical sizes and powers of the hypothesis tests for model~\eqref{eq.expam_DAR} at  $\alpha=5\%$.}
        \scriptsize
        \vspace{-1em}
	\label{tab.infer_DAR}
	\begin{center}
		\begin{tabular}{lcccccccc}
			\toprule
            & \multicolumn{2}{l}{$\btheta_0=\btheta_{H_0}$} &  \multicolumn{2}{l}{$\btheta_0=1.1\btheta_{H_0}$} & \multicolumn{2}{l}{$\btheta_0=1.3\btheta_{H_0}$} & \multicolumn{2}{l}{$\btheta_0=1.5\btheta_{H_0}$} \\
			$n$ & Wald & LM & Wald & LM & Wald & LM & Wald & LM \\
			\hline
			& \multicolumn{8}{l}{$\eta_t\sim{\rm Logistic}(0,1)$} \\
                    100   & 0.061  & 0.054  & 0.068  & 0.040  & 0.364  & 0.291  & 0.668  & 0.506  \\
    200   & 0.057  & 0.050  & 0.133  & 0.126  & 0.665  & 0.643  & 0.851  & 0.826  \\
    400   & 0.056  & 0.052  & 0.293  & 0.286  & 0.947  & 0.950  & 0.977  & 0.971  \\
          \hline

          & \multicolumn{8}{l}{$\eta_t\sim\cN(0,1.75^2)$} \\
          100   & 0.045  & 0.038  & 0.094  & 0.053  & 0.388  & 0.265  & 0.634  & 0.453  \\
    200   & 0.046  & 0.039  & 0.156  & 0.112  & 0.678  & 0.614  & 0.843  & 0.780  \\
    400   & 0.052  & 0.048  & 0.281  & 0.245  & 0.921  & 0.914  & 0.958  & 0.955  \\
          \hline

          & \multicolumn{8}{l}{$\eta_t\sim U(-2.85,2.85)$} \\
    100   & 0.058  & 0.045  & 0.131  & 0.042  & 0.381  & 0.194  & 0.595  & 0.322  \\
    200   & 0.064  & 0.061  & 0.172  & 0.093  & 0.588  & 0.448  & 0.777  & 0.626  \\
    400   & 0.044  & 0.046  & 0.287  & 0.199  & 0.873  & 0.819  & 0.889  & 0.836  \\
        \hline

        & \multicolumn{8}{l}{$\eta_t\sim 1.25 t_3$} \\
    100   & 0.058  & 0.068  & 0.048  & 0.082  & 0.298  & 0.348  & 0.616  & 0.602  \\
    200   & 0.045  & 0.046  & 0.077  & 0.155  & 0.628  & 0.763  & 0.831  & 0.921  \\
    400   & 0.044  & 0.049  & 0.218  & 0.312  & 0.954  & 0.984  & 0.962  & 0.993  \\
    
			\bottomrule
		\end{tabular}
	\end{center}
    \vspace{-2em}
\end{table}

From the tables, we can see that (i) for $\eta_t$ generated by four different random variables, the empirical sizes of the Wald test and the Lagrange multiplier test are close to the nominal significance level $\alpha=5\%$ in most cases when $n$ is large; (ii) with a larger $n$, the empirical powers of the two hypothesis tests show increasing trends. For example, under the ARMA-GARCH model~\eqref{eq.expam_AG}, for $H_1:\btheta_0=1.3\btheta_{H_0}$, $\eta_t$ is generated by $\cN(0,1.75^2)$, when $n$ increases from 100 to 400, the empirical powers of $\cT_n^{\W}$ and $\cT_n^{\LM}$ increase quickly from 0.491 to 0.975, and from 0.463 to 0.993, respectively. 
Similar simulation studies on a GARCH(1,1) model will be presented in Section~{\color{blue}S.4.2} of the supplementary material. 
In summary, our numerical results provide strong support to our theoretical results in Sections~\ref{sec.theory} and \ref{sec.application}.

\section{Real Data Analysis}
\label{sec.real}

\begin{figure}[H]
\begin{center}
\includegraphics[width=1\linewidth]{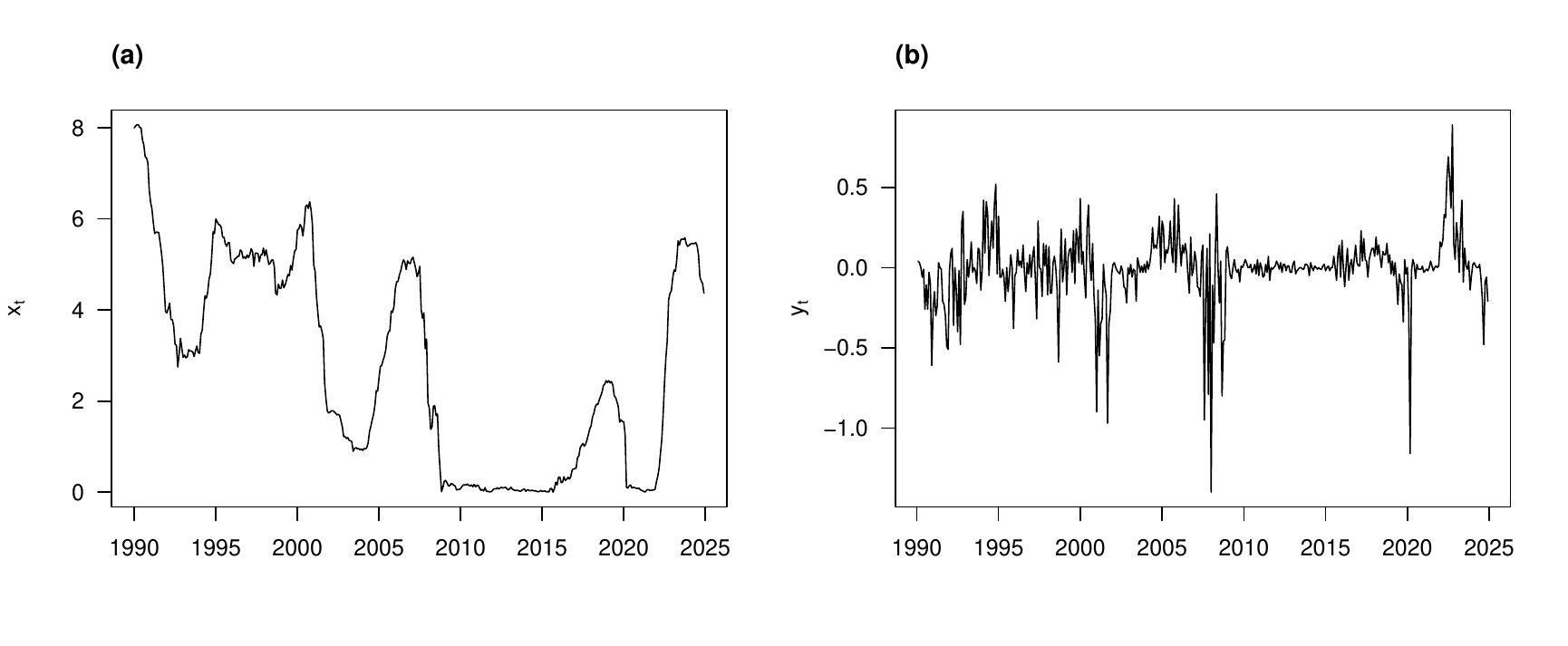}
\end{center}
\vspace{-1em}
\caption{The U.S. monthly 3-month treasury par yield curve rate series (in percent) $\{x_t\}_{t=0}^{419}$ (left panel), and its difference series $\{y_t\}_{t=1}^{419}$ (right panel).}
\vspace{-1em}
\label{fig.curve}
\end{figure}

To showcase our method, we consider the U.S. monthly 3-month treasury par yield curve rate series from January 1990 to December 2024 with a total of 420 observations, which are available at \url{https://home.treasury.gov/}. Denote the data by $x_t$ and $y_t=x_t-x_{t-1}$ as the first difference of $x_t$. The Phillips-Perron test \citep{phillips1988testing} suggests that $x_t$ is not stationary, while $y_t$ can be viewed as stationary. Figure~\ref{fig.curve} plots the monthly series $\{x_t\}_{t=0}^{419}$ and $\{y_t\}_{t=1}^{419}$, respectively. \cite{li2023maximum} also studied this dataset (until September 2021) and used an $\alpha$-stable DAR to fit the series $\{y_t\}.$ Their proposed test  indicates that the innovation follows a symmetric $\alpha$-stable distribution with $1<\alpha<2$ which corresponds to the heavy-tailed distribution without finite second-order moment. 

\begin{figure}[H]
\begin{center}
\includegraphics[width=1\linewidth]{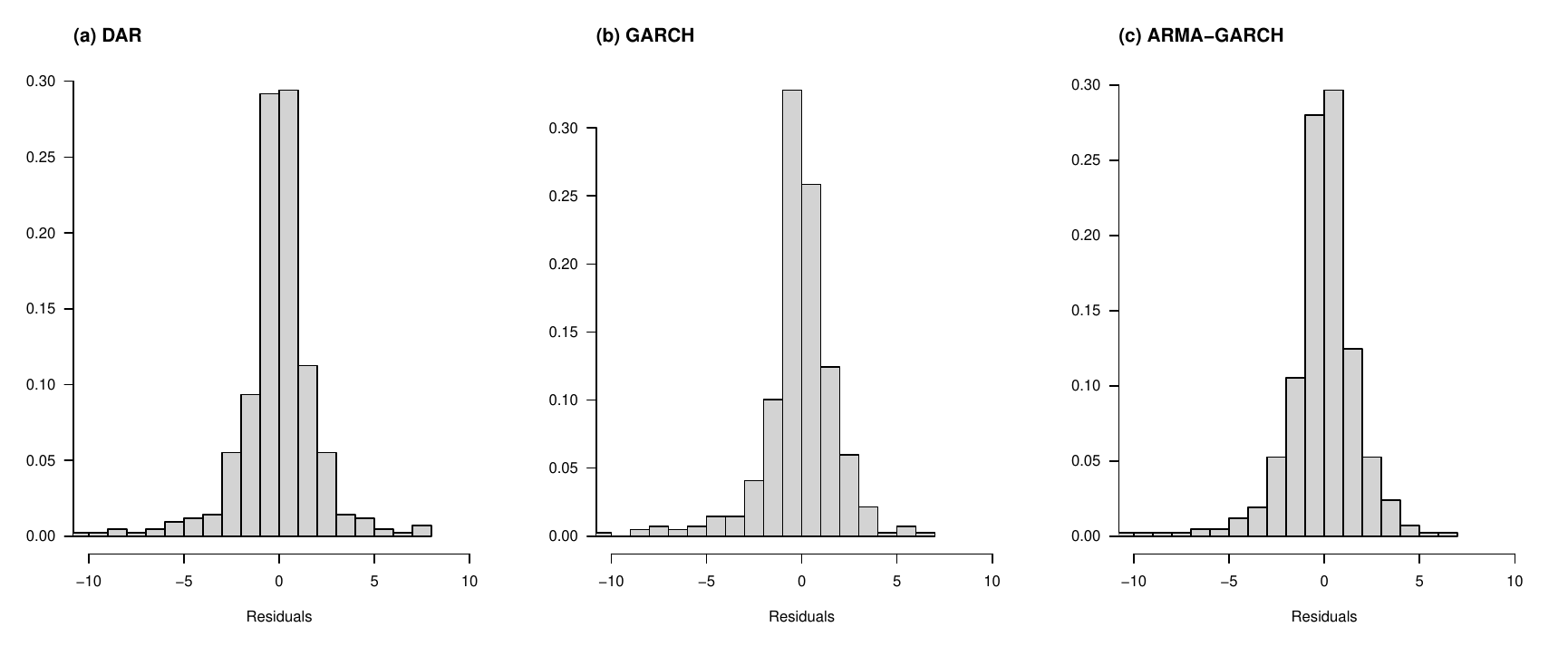}
\end{center}
\vspace{-1em}
\caption{The histograms of the residuals $\{\widehat{\eta}_t\}$ in three fitted models.}
\vspace{-1em}
\label{fig.eta}
\end{figure}

Then we fit three different models discussed in Section~\ref{sec.application}, i.e., DAR(1,1), GARCH(1,1), and ARMA(1,1)-GARCH(1,1) models, respectively, to the series $\{y_t\}_{t=1}^{419}.$ The fitting results and related statistics of all three models are 
summarized in Table~\ref{tab.fitting}, which provides the following information. First, ARMA(1,1)-GARCH(1,1) model achieves the largest log-likelihood (226.778) and the smallest AIC ($-441.555$) among these three models, and the parameters $\phi_1,\varphi_1,\alpha_0,\alpha_1$ and $\beta_1$ are significant at the significance level $\alpha=5\%$. Second, the Lyapunov exponents of these three models are $-0.6524$, $-0.7863$, and $-0.6437$, respectively, which are all less than 0 and indicate that the series is strictly stationary and ergodic. Third, the Hill estimators of the residuals $\{\widehat{\eta}_t\}$ in these three models are 0.8163, 0.6716, and 0.7377, respectively, which are all less than 1 and indicate that $\{\widehat{\eta}_t\}$ tends to be heavy-tailed. The histograms of the residuals $\{\widehat{\eta}_t\}$ in these three models are plotted in Figure~\ref{fig.eta}. Thus, it is more appropriate to model the given using our LQMLE.

\begin{table}[H]
	\caption{The fitting results and related statistics of three time series models. In the table, ``Est.'' means the parameter estimations, and ``ASD'' means the estimated asymptotic standard deviations, which are calculated by~\eqref{eq.ASD} and Theorem~\ref{thm.clt}.}
        \footnotesize
        \vspace{-1em}
	\label{tab.fitting}
	\begin{center}
		\begin{tabular}{lcccccc}
			\toprule
            & \multicolumn{2}{c}{DAR(1,1)} &  \multicolumn{2}{c}{GARCH(1,1)} & \multicolumn{2}{c}{ARMA(1,1)-GARCH(1,1)}  \\
			Parameter & Est. (ASD) & $p$-value & Est. (ASD) & $p$-value & Est. (ASD) & $p$-value \\
			\hline
                    $\phi_0$ & 0.0015 (0.005) & 0.767 & & & 0.0017 (0.001) & 0.074 \\
                    $\phi_1$ & 0.3850 (0.070) & 0.001 & & & 0.8686 (0.037) & 0.001 \\
                    $\varphi_1$ & & & & & -0.7086 (0.039) & 0.001 \\
                    $\alpha_0$ & 0.0031 (0.001) & 0.001 & 0.0010 (0.001) & 0.121 & 0.0010 (0.001) & 0.034 \\
                    $\alpha_1$ & 0.3323 (0.120) & 0.006 & 0.2867 (0.070) & 0.001 & 0.1779 (0.049) & 0.001 \\
                    $\beta_1$ & & & 0.3174 (0.102) & 0.002 & 0.4443 (0.093) & 0.001 \\
          \hline
          Lyapunov exponent & \multicolumn{2}{c}{-0.6524} & \multicolumn{2}{c}{-0.7863} & \multicolumn{2}{c}{-0.6437} \\
            Hill estimator & \multicolumn{2}{c}{0.8163} & \multicolumn{2}{c}{0.6716} & \multicolumn{2}{c}{0.7377} \\
          Log-Likelihood & \multicolumn{2}{c}{170.517} & \multicolumn{2}{c}{206.599} & \multicolumn{2}{c}{226.778} \\
          AIC & \multicolumn{2}{c}{-333.035} & \multicolumn{2}{c}{-407.199} & \multicolumn{2}{c}{-441.555} \\
			\bottomrule
		\end{tabular}
	\end{center}
        \vspace{-2em}
\end{table}

\section{Conclusion and Discussion}
\label{sec.discussion}
In this article, we have proposed a novel logistic quasi-maximum likelihood methodology in the context of time series analysis. It enjoys numerous advantages. Specifically,  it is robust in respect of distributional misspecification and heavy-tailedness of the innovation, and is more resilient to outliers than the Gaussian quasi-maximum likelihood method and the least squares method. In our asymptotic theory,  conditional symmetry of the innovation is assumed, which is mild in practice.  When there exist asymmetry phenomena in the data, we can introduce asymmetric structures or threshold effects in time series models to circumvent asymmetry issues. 

In this article, we only focus on univariate parametric time series models. There are several topics worthy of further study. For example, with our methodology, we can consider spatial autoregressive models, threshold autoregressive models with unknown threshold parameters, multivariate autoregressive models involving a joint multivariate logistic distribution as in \cite{malik1973multivariate}.  
We leave these topics for future research.

\section*{Supplementary Materials}
The supplementary material contains proofs of all main theoretical results, the technical lemmas with their proofs, the verification of assumptions, and the additional simulation results.

\section*{Disclosure Statement}
No potential conflict of interest was reported by the author(s).





\spacingset{1.0}
\bibliographystyle{dcu}
\bibliography{main}

\end{document}